\documentclass[a4paper,10pt,twocolumn]{article}

\usepackage{amsmath, amssymb}

\usepackage{amsfonts}
\usepackage{empheq}
\usepackage{verbatim}
\usepackage{graphicx}
\usepackage{float}
\usepackage{placeins}

\usepackage{tikz}
\usetikzlibrary{patterns}
\usetikzlibrary{decorations.shapes}
\usetikzlibrary{shapes.geometric}
\usetikzlibrary{shapes.arrows}

\usepackage[latin1]{inputenc}

\usepackage{cite}
\usepackage{url}

\newcommand{\ds}[1]{\displaystyle{#1}}

\def\E{{\hbox{I\kern-.2em\hbox{E}}}}

\def\argmin{\makebox{argmin}}

\newcommand{\R}{\mathbb{R}}

\newcommand{\Z}{\mathbb{Z}}

\newcommand{\boldalpha}{{\boldsymbol \alpha}}
\newcommand{\boldbeta}{{\boldsymbol \beta}}
\newcommand{\boldgamma}{{\boldsymbol \gamma}}

\newcommand{\bolddelta}{{\boldsymbol \delta}}

\newcommand{\x}{{\mathbf x}}
\newcommand{\y}{{\mathbf y}}

\newcommand{\bop}{{\mathbf b}}
\renewcommand{\d}{{\mathbf d}}

\renewcommand{\k}{{\mathbf k}}

\newcommand{\q}{{\mathbf q}}

\renewcommand{\v}{{\mathbf v}}

\renewcommand{\refeq}[1]{~(\ref{#1})}            

\newtheorem{theorem}{Theorem}
\newtheorem{lemma}[theorem]{Lemma}

\newtheorem{corollary}[theorem]{Corollary}

\newcommand{\qed}{\nobreak \ifvmode \relax \else
      \ifdim\lastskip<1.5em \hskip-\lastskip
      \hskip1.5em plus0em minus0.5em \fi \nobreak
      \vrule height0.75em width0.5em depth0.25em\fi}

\usepackage{graphicx} 
\DeclareGraphicsExtensions{.jpg,.mps,.pdf,.png} 

\usepackage{fancyhdr}
\graphicspath{
{FIGURES/}
}

\begin{document}

\title{Statistical performance analysis of a fast \\ super-resolution technique using noisy translations}

\author{Pierre Chainais,
\thanks{Pierre Chainais is with the LAGIS CNRS UMR 8219 at Ecole Centrale Lille, France. E-mail: pierre.chainais@ec-lille.fr}
Aymeric Leray,
\thanks{Aymeric Leray is with ICB, CNRS UMR 6303, Université de Bourgogne, Dijon, France. E-mail: aymeric.leray@u-bourgogne.fr}}
\date{October 23rd, 2014}

\markboth{preprint - 23 octobre 2014}{}

\maketitle

\begin{abstract}
It is well known that the registration process is a key step for super-resolution reconstruction. In this work, we propose to use a piezoelectric system
that is easily adaptable on all microscopes and telescopes for controlling accurately their motion (down to nanometers) and therefore acquiring multiple images of the same scene at different controlled positions. Then a fast super-resolution algorithm \cite{eh01} can be used for efficient super-resolution reconstruction. In this case, the optimal use of $r^2$ images for a resolution enhancement factor $r$ is generally not enough to obtain satisfying results due to the random inaccuracy of the positioning system. Thus we propose to take several images around each reference position. We study the error produced by the super-resolution algorithm due to spatial uncertainty as a function of the number of images per position. We obtain a lower bound on the number of images that is necessary to ensure a given error upper bound with probability higher than some desired confidence level.

\end{abstract}
%



%


\section{Introduction}
\label{intro}

{N}{umerous} super-resolution (SR) methods combining several low-resolution (LR) images to compute one high-resolution (HR) image have been developed and applied in microscopy, astronomy or camera photography, see \cite{petm09} for a review. However, most precise methods require long computational time. By simplifying assumptions (e.g., displacements of images are exactly known), SR methods can be faster but their range of applications may be limited. 
In this work, we propose a new cheap, fast and efficient SR technique based on a looped piezoelectric positioning stage (with precision around 0.1 nm) combined with a fast SR algorithm \cite{eh01} assuming that displacements are exactly known.
This piezoelectric stage permits to acquire multiple images of the same scene at various positions and then to assume that the registration problem is solved, at least in good approximation.
To reach a given integer resolution enhancement factor $r$ (2, 3...), we need to perform $r^2$ translations corresponding to displacements of $(k/r,\ell/r)$ in low resolution pixel units (1 LR pixel $=$ $r$ HR pixels) for integers $(k,\ell)\in(0,r-1)^2$.
In practice, the positioning system only approximately reaches the targeted positions with small random error. 
Based on a statistical performance analysis, we study the influence of this error on the quality of the SR images reconstructed with fast SR algorithm [1]. We also study the importance of using several acquisitions of the same targeted displacements on the quality of SR image in order to optimize the number of images required to ensure a given accuracy.

Over the last 30 years, several works have dealt with mathematical analysis of SR algorithms, e.g. \cite{nb02,nb03, bkz94,bk02,tla12,cbk09,rm06a,th84,ls04}. 
The works described in \cite{bkz94,nb02,nb03} essentially focus on the study of the convergence of iterative methods for superresolution (e.g., conjugate gradient) including registration and deconvolution steps. They show that the reconstruction error decreases as the inverse of the number of LR images.
In \cite{bk02}, the difficulty of the inverse problem is characterized by the conditioning number of a matrix defined from the direct model which is proportional to $r^2 s^2$ ($s=$ width of sensor pixels). When translations are uniformly distributed in $(0,r)^2$, this conditioning number tends to 1 and a direct inversion is possible with high probability when a large number of images is used \cite{tla12}. In \cite{cbk09},  the performance analysis was performed in the Fourier domain and also showed that the mean square error decreases as the number of images increases when random translations are used.
The authors of \cite{rm06a} have quantified the limitations of  superresolution methods by computing Cramer-Rao lower bounds,  also working in the Fourier domain. In the most favourable case where translations are known (no registration is needed), this bound is proportional to $r/n$ if $n$ is the number of images.

All these works and others back to the 1980s \cite{th84} explain what makes superresolution difficult and how far more images can make it simpler. However, they have only expressed limited quantitative prediction beyond the qualitative $1/n$ behaviour of the reconstruction error. Our purpose is to design a technique for which a detailed quantitative statistical error analysis can be performed. 
We obtain a lower bound on the number of images that is necessary to respect a given error bound with high probability.  
Such a control of errors is crucial not only to produce nice looking results but also to ensure reliable scientific observations.
The present study is performed in the Fourier domain and assumes that motions are exactly known (no registration) while they are only approximate and randomly distributed in practice. The precise error of each frequency component is analyzed and quantitatively evaluated. The use of Hoeffding's inequality permits to compute upper bounds so that confidence intervals of practical use are obtained. 
A classical two-step process is used : first comes the fusion of LR images by simple interlacing followed by the deconvolution step.
This method is sub-optimal to the joint treatment but leads to a simplified algorithm which has two main advantages \cite{pe09} : it is fast and the performance analysis is possible. Since the deconvolution step is conventional and well-known, we focus hereafter on the fusion step only.
A preliminary work was presented at ICASSP 2014 in \cite{clp14} based on more restricted assumptions and using the weaker Bienaymé-Cebycev inequality in place of Hoeffding's. 

Section~\ref{superresolution} presents the setting and the model. Section~\ref{errorbound} presents our main theoretical results which predict the required number of image acquisitions at each position to ensure some given confidence level in the reconstructed image. Note that sections \ref{approx} \& \ref{aliasing} present the most technical aspects and proofs while section~\ref{mainresults} sums up our main theoretical results. Section~\ref{numexp} presents numerical experiments and results. Section~\ref{conclusion} discusses our contributions and some prospects. 

\section{A fast and controlled super-resolution technique}
\label{superresolution}


\subsection{The super-resolution problem}

For a given resolution enhancement factor $r$, the most common linear formulation of the general SR problem  in the pixel domain is \cite{eh01}:
\begin{equation}
\label{SR_model}
	Y_{k}~=~D_k H_k F_k Y_{HR} + n_k \qquad k=1,..,K,
\end{equation}
where $Y_{HR}$ is the (desired) high resolution image one wants to estimate from the $K$ low-resolution images $\{Y_k, 1\leq k\leq K\}$ and $n_k$ is the noise, generally assumed to be Gaussian white noise so that $E(n_{k}n_{k}^{t}) = \sigma^2 I$. Images $Y_{HR}$, $Y_k$ and $n_k$ are rearranged in lexicographic ordered vectors. We assume the unknown high resolution image $Y_{HR}$ is a periodic bandlimited image sampled above the Nyquist rate. Each image $Y_k$ is a low resolution observation of the same underlying scene translated by $F_k$. The blur matrices $H_k$ model the point spread function (PSF) of the acquisition system and matrices $D_k$ are the decimation operator by a factor $r$. If $Y_{HR}$ is of size $r^2N^2\times 1$ and $Y_k$ of size $N\times N$, matrices $F_k$ and $H_k$ are of size $(rN)^2\times (rN)^2$ while $D_k$ are $N^2\times (rN)^2$.
The least squares optimization problem can be formulated as: 
\begin{equation}
\label{optim_pb}
	\widehat{Y_{HR}} ~=~ \argmin_{Y} \sum_{k=1}^{K} \|Y_{k}-D_{k}H_{k}F_{k} Y \|_2^2.
\end{equation}
Other formulations based on the L1-norm or adding some regularization have also been proposed \cite{frem04a}. Many methods rely on iterative optimization methods which are often computationally expensive and time consuming. For applications to biological (alive) systems are concerned, a fast method is necessary. In the present setting, displacements are controlled by a piezoelectric platform with a precision of 0.1~nm in every direction.
 The method described in \cite{eh01} appears to be a good choice since it is fast (even though more recent algorithms may yield better empirical results) and it enables to quantitatively analyze its performances and estimate error bounds with confidence level as a function of the number of available images. This is crucial for scientific imaging.
 

\subsection{Super-resolution algorithm}

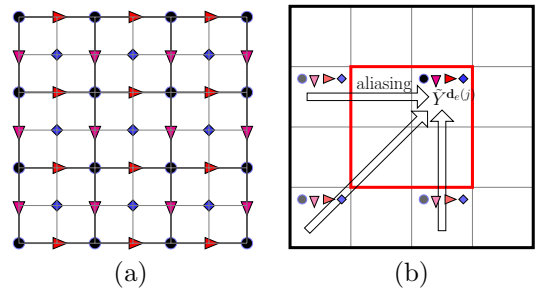
\begin{figure}
\begin{center}
\begin{tabular}{cc}
\scalebox{0.5}{
\begin{tikzpicture}		
		\foreach \x in {0,2,4,6} 
		\foreach \y in {0,2,4,6}  
			{{
			\draw (\x,\y) node [shape=circle,draw=blue!50,fill=black,minimum size=3mm,inner sep=0pt, draw] {};
			}}
		\draw[step=2cm, very thick] (0,0) grid (6,6);
		\foreach \x in {0,2,4,6} 
		\foreach \y in {1,3,5} 
			{{
			\draw (\x,\y)node[shape aspect=1,isosceles triangle,fill=magenta,rotate=-90,minimum size=0.3cm,inner sep=0pt, draw] {};
			}}
		\foreach \x in {1,3,5} 
		\foreach \y in {0,2,4,6} 
			{{
			\draw (\x,\y) node[shape aspect=2,isosceles triangle,fill=red,rotate=0,minimum size=3mm,inner sep=0pt, draw] {};
			}}
		\foreach \x in {1,3,5}
		\foreach \y in {1,3,5} 
		{{
			\draw (\x,\y) node[shape aspect=1,diamond,fill=blue!80,minimum size=3mm,inner sep=0pt, rotate=0,draw] {};			
			}}
		\draw[help lines] (0,0) grid +(6,6);
\end{tikzpicture}}
&
\scalebox{0.4}{
\begin{tikzpicture}
	\draw[step=2cm,gray,very thin] (-4cm,-4cm) grid (4cm,4cm);
	\draw[line width=1mm,red] (-2,-2) rectangle (2,2);
	\draw[line width=1mm] (-4,-4) rectangle (4,4);
	\node at (-1.5cm,1cm) [single arrow,minimum height=4cm,minimum width=3mm,draw] {};
	\node at (-1.5cm,-1.5cm) [single arrow,minimum height=5.6cm,rotate=45,draw] {};
	\node at (1cm,-1.5cm) [single arrow,minimum height=4cm,rotate=90,draw] {};

	\node at (1.4cm,1cm) {\LARGE $\tilde{Y}^{\d_e(j)}$};
	\node at (-0.9cm,1.4cm) {\LARGE aliasing};
	
	\draw (-3.6cm,1.6cm) node [shape=circle,draw=blue!50,fill=black!60,minimum size=3mm,inner sep=0pt, draw] {};
	\draw (-3.2cm,1.6cm) node[shape aspect=1,isosceles triangle,fill=magenta!60,rotate=-90,minimum size=0.3cm,inner sep=0pt, draw] {};
	\draw (-2.8cm,1.6cm) node[shape aspect=2,isosceles triangle,fill=red!60,rotate=0,minimum size=3mm,inner sep=0pt, draw] {};
	\draw (-2.3cm,1.6cm) node[shape aspect=1,diamond,fill=blue!60,minimum size=3mm,inner sep=0pt, rotate=0,draw] {};			

	\draw (-3.6cm,-2.4cm) node [shape=circle,draw=blue!50,fill=black!60,minimum size=3mm,inner sep=0pt, draw] {};
	\draw (-3.2cm,-2.4cm) node[shape aspect=1,isosceles triangle,fill=magenta!60,rotate=-90,minimum size=0.3cm,inner sep=0pt, draw] {};
	\draw (-2.8cm,-2.4cm) node[shape aspect=2,isosceles triangle,fill=red!60,rotate=0,minimum size=3mm,inner sep=0pt, draw] {};
	\draw (-2.3cm,-2.4cm) node[shape aspect=1,diamond,fill=blue!60,minimum size=3mm,inner sep=0pt, rotate=0,draw] {};			

	\draw (0.4cm,-2.4cm) node [shape=circle,draw=blue!50,fill=black!60,minimum size=3mm,inner sep=0pt, draw] {};
	\draw (0.8cm,-2.4cm) node[shape aspect=1,isosceles triangle,fill=magenta!60,rotate=-90,minimum size=0.3cm,inner sep=0pt, draw] {};
	\draw (1.2cm,-2.4cm) node[shape aspect=2,isosceles triangle,fill=red!60,rotate=0,minimum size=3mm,inner sep=0pt, draw] {};
	\draw (1.7cm,-2.4cm) node[shape aspect=1,diamond,fill=blue!60,minimum size=3mm,inner sep=0pt, rotate=0,draw] {};			

	\draw (0.4cm,1.6cm) node [shape=circle,draw=blue!50,fill=black,minimum size=3mm,inner sep=0pt, draw] {};
	\draw (0.8cm,1.6cm) node[shape aspect=1,isosceles triangle,fill=magenta,rotate=-90,minimum size=0.3cm,inner sep=0pt, draw] {};
	\draw (1.2cm,1.6cm) node[shape aspect=2,isosceles triangle,fill=red,rotate=0,minimum size=3mm,inner sep=0pt, draw] {};
	\draw (1.7cm,1.6cm) node[shape aspect=1,diamond,fill=blue!80,minimum size=3mm,inner sep=0pt, rotate=0,draw] {};		
\end{tikzpicture}
}\\
(a) & (b)
\end{tabular}

\caption{\label{LR_HR_grids} (a) Spatial domain: black disks and thick grid are the original LR sampling grid, the thin grid is the target HR grid.  Other symbols are positions of 3 translated LR images of half a LR pixel ($r=2$); (b) Fourier domain: the inner (red) square contains LR frequencies $(-N/2,N/2)^2$, the outer square is the set of HR frequencies $(-rN/2,rN/2)^2$. Arrows represent aliasing. \label{illustre_aliasing}}
\end{center}
\end{figure}

We make several usual assumptions. The PSF of the acquisition system is known and spatially homogeneous so that $\forall k, H_k=H$. Decimation is the same for all images so that $\forall k, D_k=D$ in\refeq{SR_model} \&\refeq{optim_pb}. Periodic boundary conditions make $F_k$ and $H$ circulant matrices. We will also assume that the $r^2$ possible translated images at integer multiples $(k,\ell)\in(0,r-1)^2$ of the high resolution scale are available to provide an optimal setting for super-resolution~\cite{rm06a}. Then the solution to the least-square error SR problem\refeq{optim_pb} can be decomposed in two steps and that the intermediate blurred image $Z=HY_{HR}$ can be estimated by \cite{eh01}:
\begin{equation}
\label{defZ}
	Z := HY_{HR}~=~\sum_{k=1}^{r^2} F_{k}^{t}D^{t}Y_{k} 
\end{equation}
The operation in\refeq{defZ} is equivalent to a simple interlacing of LR images, see Fig.~\ref{LR_HR_grids}.
Then the final HR image results from the deconvolution of $Z$, which can be done using  any algorithm such as Wiener or Lucy~\cite{b09}. Such an approach separates the problem of super-resolution into two steps of fusion (estimating $Z$) and deconvolution (estimating $Y_{HR}$). This work focuses on the performance analysis of the fusion step only.
Recall that high frequency terms at some $\k'$ are preserved if and only if the PSF $\tilde{H}(\k')$ is not zero. 
Some prior information might be used to reconstruct missing frequencies~\cite{rm06a}. 
 
\begin{figure}
\begin{center}
{\setlength{\tabcolsep}{0.5mm}
\begin{tabular}{cccc}
	\includegraphics[height=0.065\textwidth,width=0.065\textwidth]{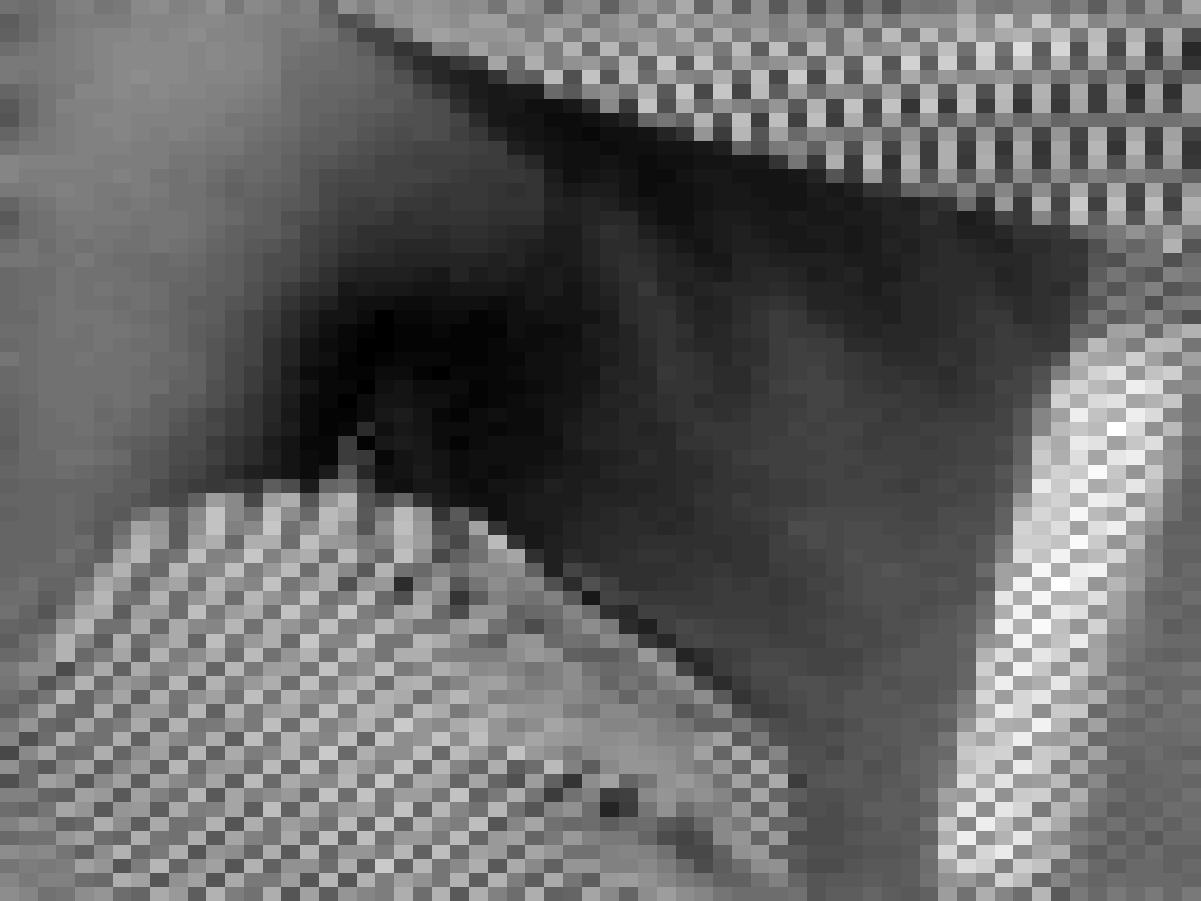}
	&
	\includegraphics[height=0.13\textwidth,width=0.13\textwidth]{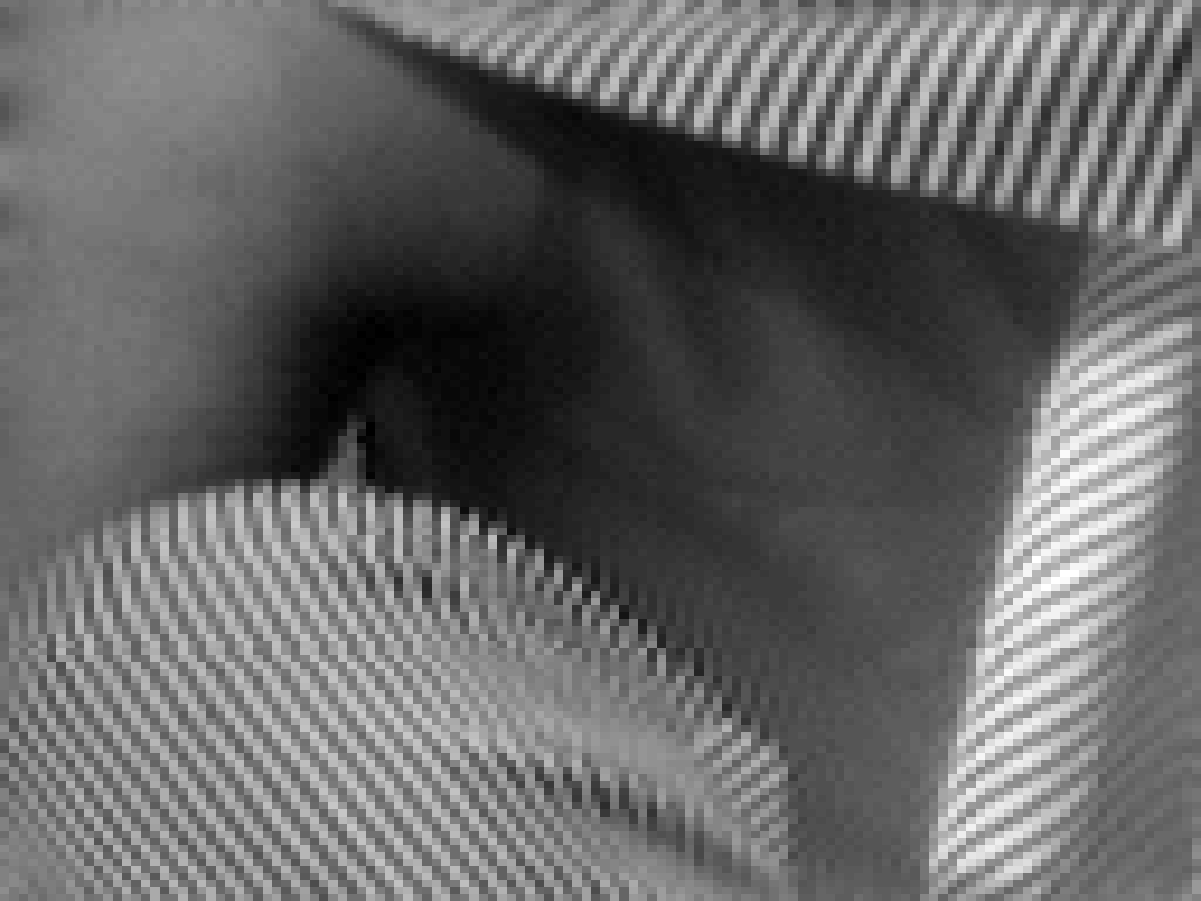}
	&
	\includegraphics[height=0.13\textwidth,width=0.13\textwidth]{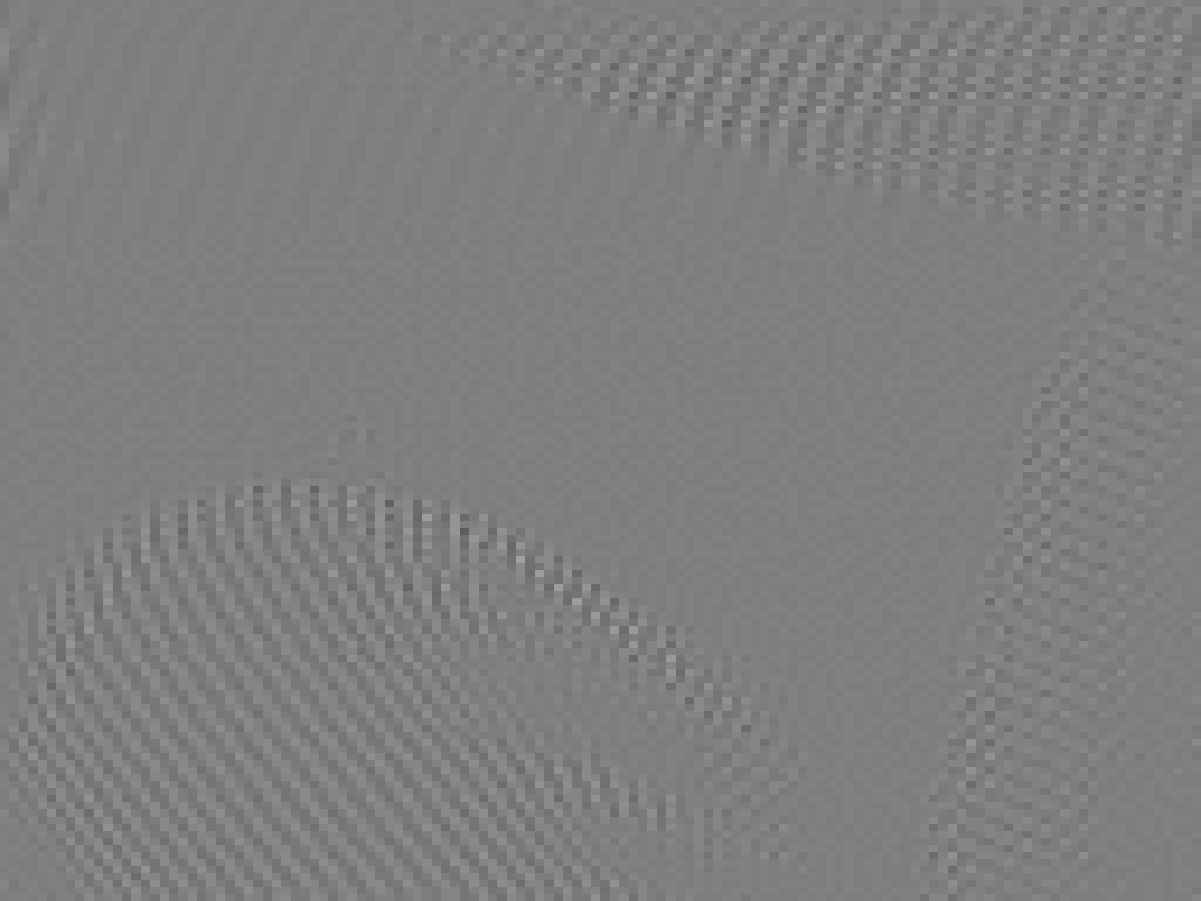}
	&
	\includegraphics[height=0.13\textwidth,width=0.13\textwidth]{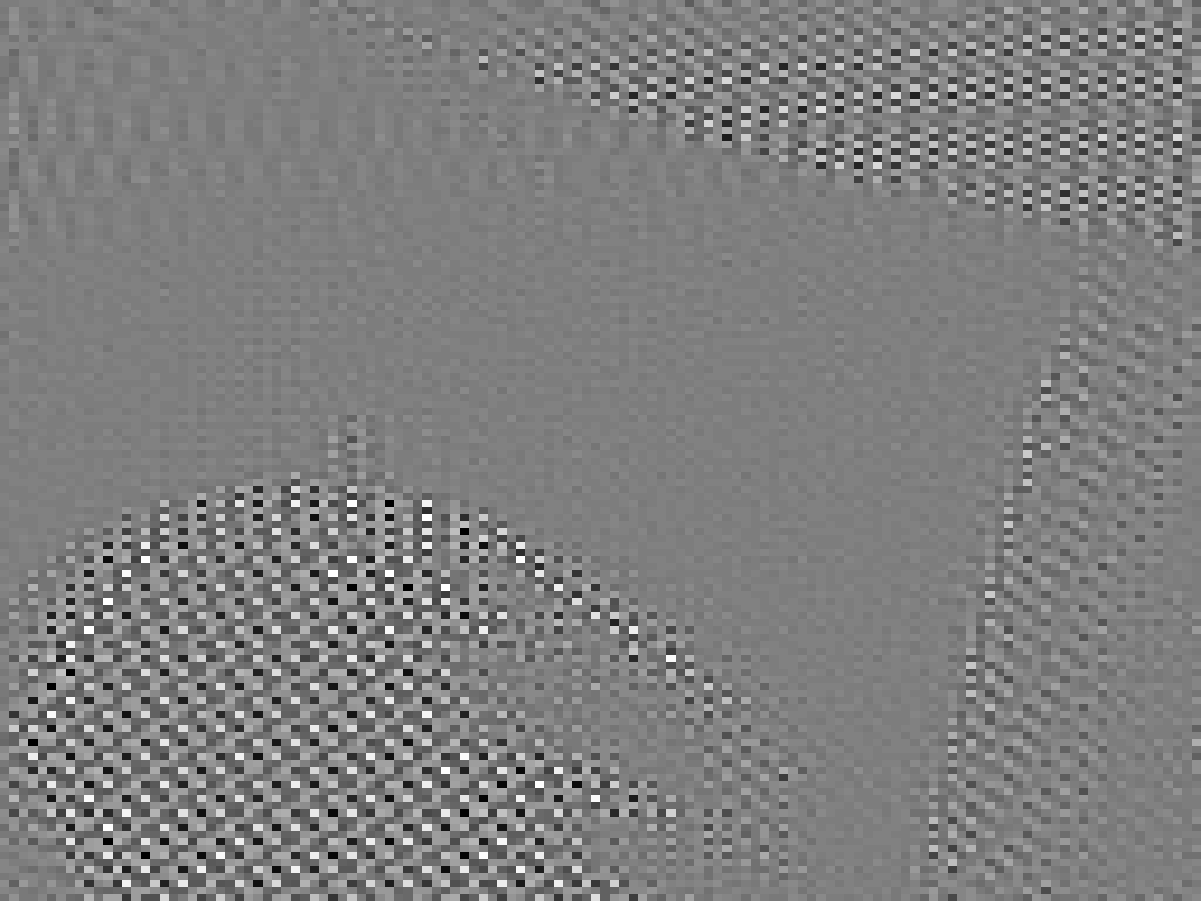}\\
	(a) & (b) & (c) & (d)
\end{tabular}
}
	\caption{\label{fig_example} (a) Barbara, (b) Results from Algorithm [1] with $r=2$, $\epsilon=0.1$, $n_\d=32$ im./pos. ; HF reconstruction error, (c) $n_d=32\Rightarrow$ SNR$_{HF} = $25.0dB, (d) $n_d=1\Rightarrow$ SNR$_{HF} = $10.0dB.}
\end{center}
\end{figure}

This algorithm is fast thanks to one idealized assumption: displacements (matrices $F_k$) are assumed to be exact integer multiples of high resolution pixels.  In practice, this assumption is only approximately true. In our setting, this is essentially due to the finite precision of the piezoelectric positioning system. One solution would be to carry out accurate sub-pixel registration but this would remain insufficient since state of the art techniques cannot ensure a precision much better than 0.1 pixel \cite{fzb02}. 
Another way to compensate positioning inaccuracies is to take $n_\d\geq 1$ images for each required position so that the true $Z$ will be replaced by the estimate: 
\begin{equation}
\label{defX_nd}
	X = \hat{Z} =  \sum_\d \frac{1}{n_\d}\sum_{j=1}^{n_\d} (F^\d)^t D^tY^{\d_e(j)}
\end{equation}
$Y^\d$ is the image of a scene $Y$ translated by $\d$ where $\d=(d_x,d_y)$ denotes the targeted displacement vector; $\d_e(j)=\d + \bop_{\d j}$ the real experimental displacement; $\bop_{\d j}$ is the noise on the platform position. Note that in general $(F^{\d})^t F^{\d_e(j)}\neq I_{rN}$. 
One can hope to compensate from displacement inaccuracies by using multiple acquisitions at the same targeted position with some random error $\bop_{\d j}$ around the expected value $\d$. An important assumption is that the position error is assumed to be bounded by $\epsilon>0$ in LR pixel units or $\epsilon_r=\epsilon r$ in HR pixel units. For a given targeted position, the positioning system will be reset between each acquisition so that positions are randomly distributed around the average position (which may be biased). This averaging process is expected to enhance the super-resolution quality. 
Fig.~\ref{fig_example} illustrates typical results from this approach applied to a detail of Barbara for $r=2$, $\epsilon=0.1$. The error on reconstructed high frequency components are compared for $n_\d=1$ and $n_d=32$ im./pos.. A gain of about 15dB is observed when using 32 images (note for later use that 10$\log_{10}(32) = 15$).
Our aim is to reconstruct {\em probably approximately correct} (PAC) images by quantifying the number of images that should be taken per reference position to respect some given upper relative error bound of $p$ (e.g. 0.10) with probability (confidence) higher than $P$ (e.g. 0.90). 

\subsection{Aliasing effects and notations}
\label{aliasingeffects}

To detail the effect of aliasing, we consider the relation between the estimated blurred HR image $X$ defined by\refeq{defX_nd} and the LR images $Y^{\d_e(j)}$ in the Fourier domain, see Fig.~\ref{illustre_aliasing}(b). For some integer $n$, the interval $(-n:n)$ denotes the set of integers between $-n$ and $n$ (Matlab notations). When using the Discrete Fourier Transform (DFT), we denote by $\k$ the LR frequencies in ${\cal D}_{LR}=(-N/2:N/2-1)^2$ and $\k'$ the HR frequencies in  ${\cal D}_{HR}=(-rN/2:rN/2-1)^2$. 
Given some HR frequency $\k'$, we need to deal with corresponding aliased terms in the LR image.
The integer vector $\boldgamma\in(-r:r)^2$ is such that $\k=\k'-\boldgamma N \in {\cal D}_{LR}$. We denote by $\boldalpha$ the integer vectors such that $\k + \boldalpha N\in {\cal D}_{HR}$. Sums $\sum_\d$ are over all the $r^2$ ideal displacements $ \d\in(0:r-1)^2$ and sums over $\boldalpha$ are sums over all possible HR frequencies $\k_\boldalpha = \k+\boldalpha N$ (up to $rN/2$). 
The DFT of image $Z$ is $\tilde{Z}$. 
To alleviate formulas, we introduce the normalized frequencies:
\begin{equation}
\left\{
\begin{array}{rcl}
\q_\boldgamma & = & \ds{\frac{2\pi}{rN}\k' = \frac{2\pi}{rN}(\k+\boldgamma N) = \q + \boldgamma \frac{2\pi}{r}}\\[2mm]
\q_\boldalpha    & = & \ds{\frac{2\pi}{rN}(\k+\boldalpha N) = \q + \boldalpha \frac{2\pi}{r}}
\end{array}
\right.
\end{equation}
where $\boldalpha,\boldgamma\in\Z^2$. Note that $\q_{\boldalpha}\in(-\pi,\pi)^2$ so that $\|\q_\boldalpha\|_1\leq 2\pi$ and $\|\q_\boldalpha\|_2\leq \sqrt{2}\pi$. 

Back to\refeq{defX_nd}, note that when $D$ is the decimation operator, $D^{t}$ is an upsampling operation (inserting zeros between samples) that produces aliasing. If ${\cal F}_{HR}$ is the HR DFT, for $\k'\in {\cal D}_{HR}$:
\begin{equation}
\label{yde}
	[{\cal F}_{HR} D^t Y^{\d_e(j)}](\k') =  \tilde{Y}^{\d_e(j)}(\k=\k'-\boldgamma N)
\end{equation}
Taking phase shifts due to translations of $(-\d)$ associated to $(F^\d)^t$ into account in the DFT of\refeq{defX_nd} yields:
\begin{eqnarray}
\label{XYde}
	\tilde{X}(\k')=\sum_\d \frac{1}{n_\d}\sum_{j=1}^{n_\d} \tilde{Y}^{\d_e(j)}(\k'-\boldgamma N)\:e^{\frac{2i\pi}{rN}\d\cdot\k'}
\end{eqnarray}
%
Since each observation is a decimated version of the blurred translated scene, one has in the spatial domain:
\begin{equation}
\label{yde_yHR}
	Y^{\d_e(j)}  = DHF^{\d_e(j)} Y_{HR} = D Z^{\d_e(j)} 
\end{equation} 
In the Fourier domain: 
\begin{equation}
\tilde{Y}^{\d_e(j)}(\k)=  [{\cal F}_{LR} D  {\cal F}_{HR}^t \tilde{Z}^{\d_e(j)}](\k)
\end{equation}
and thanks to usual properties of the sum of roots of unity (see Appendix~\ref{rootsofunity}):
\begin{equation}
\label{YYs}
	\tilde{Y}^{\d_e(j)}(\k)=\frac{1}{r^2}\sum_{\boldalpha}  \tilde{Z}(\k_\boldalpha)\: e^{-\frac{2i\pi}{rN}\k_\boldalpha \cdot \d_e(j)}
\end{equation}
where we have used the fact that the homogeneous blur operator is diagonal in Fourier domain. One can explicitly see in\refeq{YYs} how the information at high frequencies $\k_\boldalpha = \k+\boldalpha N$ from the HR image is aliased at low frequency $\k$ in each LR image $Y^{\d_e(j)}$. 
By separating the desired main contribution at $\k'=\k+\boldgamma N$ and aliasing terms at $\k+\boldalpha N$ for $\boldalpha\neq \boldgamma$, one gets by reporting\refeq{YYs} in\refeq{XYde}:
\begin{equation}
\label{X_decomp2}
\tilde{X}(\k')=\tilde{Z}(\k')G_{\boldgamma}(\k') + \underbrace{\sum_{\boldalpha\neq\boldgamma}\tilde{Z}(\k+\boldalpha N) G_{\boldalpha}(\k')}_{B(\k')}
\end{equation}
where 
\begin{equation}
\label{def_Galpha}
	G_{\boldalpha}(\k')	=  \frac{1}{r^2 n_\d}\sum_{\d,j} e^{-i\frac{2\pi}{r}(\boldalpha-\boldgamma)\d} e^{i\q_\boldalpha\cdot \bop_{\d j}} 
\end{equation}
(except when $k'_x$ or $k'_y$ is equal to $-rN/2$). 
 In the ideal case where $\bop_{\d j}={\mathbf 0}$ so that translations are exact multiples of HR pixels, one retrieves $\tilde{X}=\tilde{Z} = \tilde{H}\tilde{Y}_{HR}$ since $G_\boldgamma = 1$ and $G_\boldalpha = 0$ for $\boldalpha\neq\boldgamma$.
The first term in\refeq{X_decomp2} is the {\em main approximation term}, which should be as close as possible to $\tilde{Z}(\k')$. The second term $B(\k')$ is the {\em aliasing term} and should be as small as possible compared to the approximation term. Our purpose is to establish conditions for which $X$ is a good approximation of $Z$ within quantitative probabilistic bounds. 
Note that under {\em the perfect deconvolution assumption} with $\tilde{H}(\k')\neq 0$ everywhere one could estimate $Y_{HR}$ from $X$ by using $\tilde{Y}(\k') = \tilde{X}(\k')/\tilde{H}(\k')$, all results on relative errors on $Z$ directly apply to $Y$. This is not true in general since the deblurring restoration step will introduce some errors ; however the present analysis of the fusion step still holds.

\section{Probabilistic bounds on reconstruction errors}
\label{errorbound}

The purpose of this section is to obtain concentration inequalities that guarantee {\em PAC} superresolution. In this study, position errors $\bop_{\d j}\in(-\epsilon_r,\epsilon_r)^2$ HR pixel units. We do not assume that $\E[\bop_{\d j}] = 0$: the positioning system might be biased. In section \ref{approx} \& \ref{aliasing}  we deal with the coefficient $G_\boldgamma$ of $Z$ in the main approximation term of\refeq{X_decomp2} and then turn to the contribution of the aliasing term $B(\k')$. The reader interested in our main results only can directly move to sections \ref{mainresults} \& \ref{control_SNR}. 

\subsection{Bound on the approximation term $G_\boldgamma(\k')$}
\label{approx}
Since one expects that $\frac{1}{r^2}\E G_\boldgamma \simeq 1$, we start from
\begin{eqnarray}
	& \left| G_{\boldgamma}(\k') - 1 \right| \leq \label{decompGgamma}\\
	&   \left|  G_{\boldgamma}(\k') -  \E[G_{\boldgamma}(\k')]\right| + \left| \E[G_{\boldgamma}(\k')] - 1\right| \nonumber
\end{eqnarray}
Noting that $\E[G_{\boldgamma}(\k')] =  \E [e^{i\q_\boldgamma\cdot\bop_{\d j}}]$, the Taylor development of the complex exponential function yields\footnote{See Lemma 1 p. 512 in Feller (vol. 2) \cite{feller2} on the Taylor development of $\exp(it)$ for $t>0$.}
\begin{multline}
	\left| \E [e^{i\q_\boldgamma\cdot\bop_{\d_j}}] - 1\right| 
	\leq \ds{|\q_\boldgamma. \E[\bop_{\d j}]| + \E \left[ \frac{(\q_\boldgamma.\bop_{\d_j})^2}{2} \right]} \\
	\leq \ds{\underbrace{\|\q_\boldgamma\|_2\, \|\E[\bop_{\d j}]\|_2 + \frac{\|\q_\boldgamma\|_1^2\:\epsilon_r^2}{2}}_{B1}} \label{approx1}
\end{multline}
since $\bop_{\d j}\in(-\epsilon_r,\epsilon_r)^2$. 
Then we deal with the first term in\refeq{decompGgamma} by introducing:
\begin{equation}
\label{defBG}
	B_G = 
	\frac{1}{r^2 n_\d} \sum_{\d,j}  (e^{i\q_\boldgamma\cdot\bop_{\d_j}} -\E e^{i\q_\boldgamma\cdot\bop_{\d j}})
\end{equation}
To obtain concentration inequalities on $|B_G|$, our approach goes in 3 steps: i) bound the real and imaginary parts thanks to properties of their power series expansions, ii) prove concentration inequalities by using Hoeffding's inequality for the sum of differences between random variables and their expectations, iii) bound $|B_G|$ by using Lemma~\ref{lemma1} below to combine bounds on the real and imaginary parts.


\begin{lemma}
\label{lemma1}
(see proof in Appendix~\ref{prooflemmas}) Let $x_1$ and $x_2$ two random variables in $\R$.
Let $a_1,a_2>0$ and $P_1,P_2\in(0,1)$ such that $P(|x_i|\geq a_i)\leq P_i$, $i=1,2$.
Then
\begin{equation}
\label{sum_square}
	P(\sqrt{x_1^2+x_2^2}\geq \sqrt{a_1^2+a_2^2})\leq P_1 +  P_2
\end{equation}
\begin{equation}
\label{sum_module}
	P(|x_1|+|x_2|\geq a_1+a_2) \leq P_1 +  P_2
\end{equation}

\end{lemma}

\rm
As far as the real part of $B_G$ is concerned:
\begin{equation}
	{\cal R}e \left( B_G \right)
	  = \frac{1}{r^2 n_\d}\sum_{\d,j} \cos\left(\q_\boldgamma. \bop_{\d j}\right) - \E[\cos\left(\q_\boldgamma. \bop_{\d j}\right)]
\end{equation}
The alternating power series development of the $\cos$ function yields:
\begin{equation}
	\label{devlim_cos_a}
	\left|\cos(\q_\boldgamma\cdot \bop_{\d j}) - 1  +\frac{(\q_\boldgamma\cdot\bop_{\d j})^2}{2}\right| \leq \frac{\|\q_\boldgamma\|_1^4\epsilon_r^4}{24}
\end{equation}
so that
\begin{multline}
\label{realBG}
	\hspace*{-3mm} \left| {\cal R}e \left( B_G \right)  \right| 
	\leq 
	\left| \ds{\frac{1}{2r^2 n_\d}} \sum_{\d,j}   (\q_\boldgamma.\bop_{\d j})^2 - \E[(\q_\boldgamma.\bop_{\d j})^2]\right| \\
	\ds{+ \frac{\|\q_\boldgamma\|_1^4\epsilon_r^4}{12}} \qquad\qquad
\end{multline}
One needs to bound the first term which consists of the sum of bounded centered random variables
since both  $(\q_\boldgamma.\bop_{\d j})^2$ and $\E [(\q_\boldgamma.\bop_{\d j})^2]$ belong to $(0, \|\q_\boldgamma\|_1^2\: \epsilon_r^2)$.
Now let us recall Hoeffding's inequality.

{\em Hoeffding's inequality \cite{blm13}.} 
Let $\{X_i, 1\leq i\leq n\}$ a set of independent random variables distributed over finite intervals $[a_i,b_i]$. Let $S=\sum_{i=1}^n \left(X_i-\E[X_i]\right)$. For all $t>0$,  
\begin{equation}
\label{Hoeffding_abs}
	P(|S|\geq t) \leq 2\exp\left( - \frac{2 t^2}{\sum_{i=1}^n (b_i-a_i)^2} \right)
\end{equation}
Applying Hoeffding's inequality to the first term of\refeq{realBG} for random variables in $(-\|\q_\boldgamma\|_1^2\: \epsilon_r^2,\|\q_\boldgamma\|_1^2\: \epsilon_r^2)$ and $t=\delta_\boldgamma = c(\q_\boldgamma)\|\q_\boldgamma\|_1\epsilon$ yields:
\begin{equation}
\label{ineq_R_app}
	P\left(\left| {\cal R}e \left(B_G\right)\right|\geq \delta_\boldgamma +   \frac{\|\q_\boldgamma\|_1^4\epsilon_r^4}{12} \right) 
	\leq
	2 e^{-\frac{c^2n_\d}{2\|\q_\boldgamma\|_1^2\epsilon_r^2}}
\end{equation}
since $\sum_{\d,j}$ involves $r^2n_\d$ terms and $\epsilon_r=\epsilon r$. This is the desired concentration inequality for the real part.
Turning to the imaginary part,
\begin{equation}
	{\cal I}m \left( B_G \right)
	  =  \frac{1}{r^2 n_\d}\sum_{\d,j} \sin\left(\q_\boldgamma\cdot \bop_{\d j}\right) - \E[\sin\left(\q_\boldgamma\cdot \bop_{\d j}\right)]
\end{equation}
The alternating power series development of the $\sin$ function yields:
\begin{equation}
	\label{devlim_sin_a}
	\left|\sin(\q_\boldgamma\cdot \bop_{\d j}) - \q_\boldgamma\cdot\bop_{\d j}\right| \leq \frac{\|\q_\boldgamma\|_1^3\epsilon_r^3}{6}
\end{equation}
so that
\begin{multline}
\label{imBG}
	\left| {\cal I}m \left( B_G \right)  \right| 
	\leq \\
	\left|\frac{1}{r^2 n_\d}\sum_{\d,j} \left( \q_\boldgamma\cdot\bop_{\d j} - \E[\q_\boldgamma\cdot\bop_{\d j}]\right)\right| 
	+ \frac{\|\q_\boldgamma\|_1^3\epsilon_r^3}{3}
\end{multline}
Following the same lines as for the real part, we prove a concentration inequality similar to\refeq{ineq_R_app} by applying Hoeffding's inequality to the first term of\refeq{imBG} since 
	$\left| \q\cdot\bop - \E [\q\cdot\bop]\right| \leq  2 \|\q\|_1\: \epsilon_r$.
For $\delta_\boldgamma = c(\q_\boldgamma)\|\q_\boldgamma\|_1\epsilon>0$,
\begin{equation}
\label{ineq_I_app}
	P\left(\left| {\cal I}m \left(B_G\right)\right|\geq \delta_\boldgamma +   \frac{\|\q_\boldgamma\|_1^3\epsilon_r^3}{3} \right) 
	\leq
	2 e^{-c^2n_\d/8}	
\end{equation}
Using Lemma~\ref{lemma1} to combine\refeq{ineq_R_app} and\refeq{ineq_I_app} yields: 
\begin{multline}
	P\left(\left| B_G\right|\geq  \sqrt{ \left(\delta_\boldgamma +   \frac{\|\q_\boldgamma\|_1^4\epsilon_r^4}{12}\right)^2 + \left(\delta_\boldgamma +   \frac{\|\q_\boldgamma\|_1^3\epsilon_r^3}{3}\right)^2 } \right) \\
	\leq
	2\left( 
	e^{-c^2n_\d/8} + e^{-\frac{c^2n_\d}{2\|\q_\boldgamma\|_1^2\epsilon_r^2}}
	\right)
\end{multline}
To simplify this expression, note that $\ds{\frac{\|\q_\boldgamma\|_1^4\epsilon_r^4}{12}\leq \frac{\|\q_\boldgamma\|_1^3\epsilon_r^3}{3}}$ as soon as $\epsilon_r\leq 4/\pi$. 
Remark also that $2\|\q_\boldgamma\|_1^2\epsilon^2\leq 8\pi^2\epsilon_r^2 <8$ as soon as $\epsilon<1/\pi r$.
As a consequence:
\begin{equation}
\label{approx_conc}
	P\left( \left| B_G\right| \geq  \underbrace{\sqrt{2}\left(\delta_\boldgamma +   \frac{\|\q_\boldgamma\|_1^3\epsilon_r^3}{3}\right)}_{B2} \right) 
	\leq
	4 e^{-c^2n_\d/8}
\end{equation}
We obtain the final concentration inequality for the main approximation term by combining\refeq{approx1} and\refeq{approx_conc} and going back to\refeq{decompGgamma}:
\begin{equation}
\label{approx_final}
	P\left(\left|  G_{\boldgamma}(\k') - 1 \right|
	\geq B_1 + B_2
	\right) 
	\leq 
	4 e^{-c^2n_\d/8}
\end{equation}
for $\epsilon_r\leq 1/\pi r$, where $B_1$ and $B_2$ are defined in\refeq{approx1} \&\refeq{approx_conc}. 
%
Let $p\in(0,1)$ the maximum relative error constraint, e.g., $p=0.1$, and $P_1\in(0,1)$ such that $1-P_1$ is the corresponding concentration probability. 
For sufficiently large $p$, one can define for each $\k'\in{\cal D}_{HR}$ or $\q_\boldgamma\in \frac{2\pi}{rN}{\cal D}_{HR}$ the adequate maximum coefficient $c(\q_\boldgamma)>0$ such that, neglecting the cubic term,
\begin{equation}
\label{def_cqgamma}
	\sqrt{2}c(\q_\boldgamma)\|\q_\boldgamma\|_1\epsilon 
	+ \|\q_\boldgamma\|_2\langle \|\E[\bop_{\d,j}]\|_2\rangle_\d 
	+ \frac{\|\q_\boldgamma\|_1^2\epsilon_r^2}{2}
	\leq
	p
\end{equation}
$c(\q_\boldgamma)$ is a decreasing function of $\|\q_\boldgamma\|_1$, which is minimum for maximal frequencies such that $\|\q_\boldgamma\|_1=2\pi$. For $p$ large enough, one can define
\begin{eqnarray}
	c_1(p) = \min_{\q_\boldgamma} c(\q_\boldgamma) = c(\pi,\pi) \qquad\qquad \\
	\quad = \frac{1}{2\sqrt{2}\pi\epsilon} \left( p - \sqrt{2}\pi\langle \|\E[\bop_{\d,j}]\|_2\rangle_\d - 2\pi^2\epsilon^2 r^2\right) \nonumber
	\end{eqnarray}
Then\refeq{def_cqgamma} with $c(\q_\boldgamma)$ replaced by $c_1(p)$ is true for all $\q_\boldgamma\in \frac{2\pi}{rN}{\cal D}_{HR}$. When the averaged bias $\langle \E [e^{i\q_\boldgamma\cdot\bop_{\d j}}]\rangle_\d$ is zero or remains negligible ($\ll p/\sqrt{2}\pi$), 
\begin{equation}
	c_1(p)\simeq \frac{p-2\pi^2\epsilon^2 r^2}{2\sqrt{2}\pi\epsilon}
\end{equation}
If $\epsilon\leq 1/\pi r$ and $c_1(p)$ is well defined,\refeq{approx_final} becomes for all $\k'\in{\cal D}_{HR}$:
 \begin{equation}
	P\left(\left|  G_{\boldgamma}(\k') - 1 \right|
	\geq
	p
	\right) 
	\leq
	4 \exp\left( - \frac{c_1(p)^2 n_\d}{8}\right)
\end{equation}
Then one can guarantee that the relative error remains upper bounded by $p$ with probability larger than $P_1$ as soon as
\begin{equation}
	4\exp\left( - \frac{c_1(p)^2 n_\d}{8}\right) \leq 1-P_1
\end{equation}
which provides a first lower bound on $n_\d$
\begin{equation}
	n_\d \geq \frac{8}{c_1(p)^2}\log \left(\frac{4}{1-P_1}\right)
\end{equation}
The larger $c_1(p)$, the smaller the lower bound. This bound does not depend on the image content. 
In practice, the number $n_\d$ of images per positions must obey this bound to ensure that the main approximation term in\refeq{X_decomp2} be less than $100p$\% away from the targeted $\tilde{Z}(\k')$ with probability larger than $P_1$. 
In ideal experimental conditions, with no bias and $\epsilon r\leq \sqrt{p/2\pi^2}$, 
\begin{equation}
	n_\d \geq \left( \frac{8\pi\epsilon}{p-2\pi^2\epsilon^2 r^2}\right)^2 \log \left(\frac{4}{1-P_1}\right)
\end{equation}
For instance, see Tab.~\ref{table1}, for $\epsilon=0.01$, $r=2$, $p=0.1$ and $P_1=0.90$ (error $\leq 10$\% with $\geq 90$\% confidence level) this bound is $n_\d\geq 28$. The concentration level $(1-P_1)$ can be very tight due to the logarithmic dependence of $n_\d$ on $(1-P_1)$. At the same error level $p=0.1$, the criterion becomes $n_\d\geq 45$ for $P_1=0.99$. In contrast, a much larger $n_\d\geq 5.3\,10^4$ is necessary to guarantee an accuracy of 1\% ($p=0.01$) at $P_1=0.90$ confidence level. 
Note that the position accuracy $\epsilon$ should essentially decrease proportionally to $p$ as a finer reconstruction is desired.  
Moreover, given a desired superresolution factor $r$ and a position accuracy $\epsilon$, the relative error $p$ is lower bounded by $2\pi^2\epsilon^2 r^2$. For instance, for $r=2$ and $\epsilon=0.01$, the smallest relative error $p$ that can be guaranteed is $p_{best}=0.008$.
 


\subsection{Bound on the aliasing terms ($G_\boldalpha$, $\boldalpha\neq\boldgamma$)}
\label{aliasing}

The ideal situation in\refeq{X_decomp2} occurs when the translations $\d$ are exactly the $r^2$ possible multiples of HR pixels. Due to properties of complex roots of unity, all the aliasing terms $G_{\boldalpha}(\k')$ in\refeq{X_decomp2} are zero for $\boldalpha\neq\boldgamma$. Now we study these terms when translations are only approximately controlled. Our aim is to bound the contribution of aliased terms. The adopted strategy is similar to that of previous section by applying Hoeffding's inequality. We also use the properties of roots of unity and a standard assumption on the spectral content of the target image. 
We start from\refeq{def_Galpha}: 
\begin{equation}
G_{\boldalpha}(\k')
	 =  \frac{1}{r^2n_\d}\sum_{\d,j} e^{-i\frac{2\pi}{r}(\boldalpha-\boldgamma)\d}  e^{i\q_\boldalpha\cdot \bop_{\d j}}
\end{equation}
Let
\begin{equation}
	\theta_{\boldalpha \d} = \frac{2\pi}{r}(\boldalpha-\boldgamma)\d, \quad \d\in(0:r-1)^2
\end{equation}
Note that the set of the $e^{i\theta_{\boldalpha\d}}$ matches the set of products of complex roots of unity, see eq.\refeq{sumzerocos_app}-(\ref{sumsin2_app}) in Appendix~\ref{rootsofunity}. The sum over translations $\sum_\d$ actually involves the sum of roots of unity, which is zero, in the computation of the aliasing term.
First, we deal with the real part of $G_\boldalpha(\k')$:
\begin{equation}
	{\cal R}e \left(G_\boldalpha(\k')\right) 
	  = \frac{1}{r^2 n_\d}\sum_{\d,j} \cos\left(\theta_{\boldalpha \d}-\q_\boldalpha\cdot \bop_{\d j}\right) %
\end{equation}
The power series development of the $\cos$ function around $\theta_{\boldalpha \d}$ yields:
\begin{multline}
	\hspace*{-3mm} \cos(\theta_{\boldalpha \d}-\q_\boldalpha\cdot \bop_{\d j}) - \cos\left(\theta_{\boldalpha \d}\right)  - \sin\left(\theta_{\boldalpha \d}\right)(\q_\boldalpha\cdot \bop_{\d j})
	= \\
	\cos(\theta_{\boldalpha \d}) \sum_{k=1}^\infty (-1)^k \frac{(\q_\boldalpha\cdot \bop_{\d j})^{2k}}{(2k)!}\\
	+
	\sin(\theta_{\boldalpha \d}) \sum_{k=1}^\infty (-1)^{k} \frac{(\q_\boldalpha\cdot \bop_{\d j})^{2k+1}}{(2k+1)!}
\end{multline}
The classical theorem of majorization of the rest of alternating power series yields:
\begin{multline}
\label{devlim_cos}
	\hspace*{-4mm} \left|\cos(\theta_{\boldalpha \d}-\q_\boldalpha. \bop_{\d j}) - \cos\theta_{\boldalpha \d}  - (\q_\boldalpha. \bop_{\d j})\sin\theta_{\boldalpha \d}\right| \\
	\leq
	|\cos\theta_{\boldalpha \d}| \frac{|\q_\boldalpha\cdot \bop_{\d j}|^2}{2}
	+
	|\sin\theta_{\boldalpha \d}| \frac{|\q_\boldalpha\cdot \bop_{\d j}|^3}{6}	
\end{multline}
As a consequence,
\begin{multline}
\label{ineq_alias_inter}
	{\cal R}e \left(G_\boldalpha(\k')\right) 
	\leq  f(\q_\boldalpha,\epsilon_r) \\
	+
	\left|
	\frac{1}{r^2 n_\d}\sum_{\d,j} \cos\theta_{\boldalpha\d}
	+ (\q_\boldalpha\cdot \bop_{\d j} )\sin\theta_{\boldalpha \d}\right|
\end{multline}
where  
\begin{equation}
\label{def_f}
	f(\q_\boldalpha,\epsilon_r) = \frac{\|\q_\boldalpha\|_1^2\epsilon_r^2}{2} +  \frac{\|\q_\boldalpha\|_1^3 \epsilon_r^3}{6}
\end{equation}
Thanks to\refeq{sumzero_app} in Appendix~\ref{rootsofunity},
\begin{equation}
\label{ineq_R}
	{\cal R}e \left(G_\boldalpha\right) 
	\leq 
	\left|\frac{1}{r^2 n_\d}\sum_{\d,j} (\q_\boldalpha\cdot \bop_{\d j} )\sin\theta_{\boldalpha \d}\right| + f(\q_\boldalpha,\epsilon_r) 
\end{equation}
For $\boldalpha-\boldgamma\in \{ 0,r/2\}^2$, $\theta_{\boldalpha \d}\propto \pi \Rightarrow \sin\theta_{\boldalpha \d}=0$ for all $\d$ and\refeq{devlim_cos} yields the deterministic tight inequality
\begin{equation}
	\label{hoeffding_real_r2}
	{\cal R}e \left(G_\boldalpha(\k')\right) 
	\leq \frac{\|\q_\boldalpha\|_1^2\epsilon_r^2}{2}
\end{equation}
Turning to the imaginary part, we follow the same lines by mainly replacing '$\cos$' by '$\sin$' in\refeq{ineq_alias_inter} \&\refeq{ineq_R} starting from
\begin{multline}
\label{devlim_sin}
	\hspace*{-3mm} \left|\sin(\theta_{\boldalpha \d}-\q_\boldalpha. \bop_{\d j}) - \sin\theta_{\boldalpha \d}  + (\q_\boldalpha. \bop_{\d j})\cos\theta_{\boldalpha \d}\right| \\
	\leq
	|\sin\theta_{\boldalpha \d}| \frac{|\q_\boldalpha. \bop_{\d j}|^2}{2}
	+
	|\cos\theta_{\boldalpha \d}| \frac{|\q_\boldalpha. \bop_{\d j}|^3}{6}	
\end{multline}
to obtain the following bound on the imaginary part:
\begin{equation}
\label{ineq_I}
	\left| {\cal I}m \left(G_\boldalpha\right) \right| \leq 
	\left|\frac{1}{r^2 n_\d}\sum_{\d,j} (\q_\boldalpha. \bop_{\d j} )\cos\theta_{\boldalpha \d}\right| 
	+ f(\q_\boldalpha,\epsilon_r) 
\end{equation}
Then one needs to bound the sums in the r.h.s. of\refeq{ineq_R} \&\refeq{ineq_I}. 
Assuming that the variations of the bias $\E[\q_\boldalpha \cdot \bop_{\d j}]$ around  $\langle\E[\q_\boldalpha \cdot \bop_{\d j}]\rangle_\d$ for fixed $\d$ are negligible, one observes that there is (approximately) no contribution of the bias in the sums of\refeq{ineq_R} \&\refeq{ineq_I} thanks to\refeq{sumzerocos_app} \&\refeq{sumzerosin_app}. Indeed, 
\begin{equation}
\frac{1}{r^2 n_\d}\sum_{\d,j} 
\cos(\theta_{\boldalpha \d})\langle\E[\q_\boldalpha . \bop_{\d j}]\rangle_\d ) 
\simeq 0
\end{equation}
Since $\q_\boldalpha\cdot\bop_{\d j}\in(-\|\q_\boldalpha\|_1\epsilon_r,\|\q_\boldalpha\|_1\epsilon_r)$,  we apply Hoeffding's inequality to\refeq{ineq_R} and\refeq{ineq_I} for $\delta_\boldalpha=c\|\q_\boldalpha\|_1\epsilon$ as before, and for $\boldalpha-\boldgamma\notin \{ 0,r/2\}^2$ to obtain :
\begin{eqnarray}
	P\left(\left| {\cal R}e \left(G_\boldalpha(\k')\right)\right|
	\geq \delta_\boldalpha'
	\right) 
	\leq 
	2e^{-\frac{2 r^2 n_\d^2 c^2}{4\sum_{\d,j} \sin^2\theta_{\boldalpha \d}}}\\
	P\left(\left| {\cal I}m \left(G_\boldalpha(\k')\right)\right|
	\geq \delta_\boldalpha'
	\right) 
	\leq 
	2e^{-\frac{2 r^2 n_\d^2 c^2}{4\sum_{\d,j} \cos^2\theta_{\boldalpha \d}}}
\end{eqnarray}
where $\delta_\boldalpha' = c\|\q_\boldalpha\|_1\epsilon + f(\q_\boldalpha,\epsilon_r)$. 
Then thanks to\refeq{sumcos2_app} and\refeq{sumsin2_app} in Appendix~\ref{rootsofunity}, one gets from Lemma 1 the following concentration inequality for $\boldalpha-\boldgamma\notin \{ 0,r/2\}^2$ : 
\begin{equation}
\label{Galpha_concentration}
	P\left(
	\left| G_\boldalpha(\k')\right|
	\geq 
	\sqrt{2} \delta_\boldalpha'
	 \right) 
	\leq 
	4 e^{- c^2 n_\d}
\end{equation}
For $\boldalpha-\boldgamma\in \{ 0,r/2\}^2$, \refeq{hoeffding_real_r2} gives a deterministic bound on the real part. Moreover $\sin(\theta_{\boldalpha\d})=0$ in\refeq{devlim_sin} so that one gets from\refeq{sum_module} in Lemma~\ref{lemma1}:
\begin{equation}
\label{Galpha_concentration_r2}
	P\left( 
	\left| G_\boldalpha(\k')\right|
	\geq \delta_\boldalpha'
	\right) 
	 \leq 
	2 e^{- c^2 n_\d}
\end{equation}
which is even tighter than\refeq{Galpha_concentration}.  In the special case $r=2$, all $\boldalpha-\boldgamma$ are in $\{ 0,r/2\}^2=\{ 0,1\}^2$ so that we need \refeq{Galpha_concentration_r2} only and tighter bounds are obtained. 

We aim at taking into account the contribution of all terms $\tilde{Z}_\boldalpha G_\boldalpha (\k')$ for $\boldalpha\neq\boldgamma$ in \refeq{X_decomp2}. Let assume that they are independent. This is at least approximately true for two main reasons. First one can show that the $G_\boldalpha (\k')$ are uncorrelated, see\refeq{Guncorrelated} in Appendix~\ref{Guncor} and second the $\tilde{Z}_\boldalpha$ carry information about very distinct frequencies in the image. Then we can use Lemma~\ref{lemma3} (see proof in Appendix~\ref{prooflemmas}):
\begin{lemma}
\label{lemma3} 
Let $x_i$, $i=1,...,n$, $n$ independent random variables.
Let $a_i>0$ and $P_i\in(0,1)$  $i=1,...,n$, such that 
$\forall i, P(|x_i|\geq a_i)\leq P_i$.
Then
\begin{equation}
	P\left(\sum_i |x_i| \leq \sum_i a_i\right) \geq \prod_{i=1}^n (1-P_i)
\end{equation}
\end{lemma}

\begin{figure}
\begin{center}
	\includegraphics[height = 0.3\textwidth,width=0.4\textwidth]{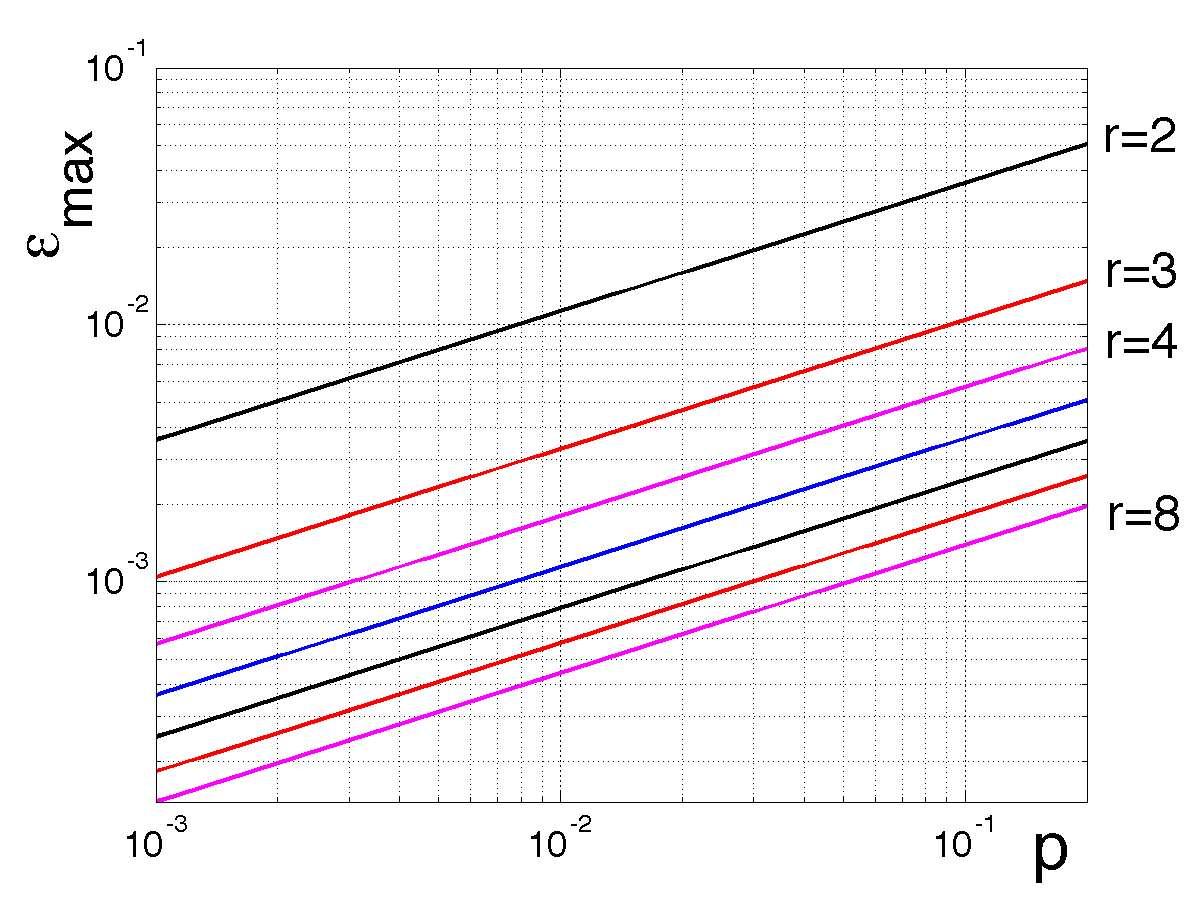}
	\caption{\label{eps_p_bound} Pairs of parameters $(p,\epsilon)$ for which superresolution with guaranted error bounds is feasible are at the bottom right of the curve for each $r$.}
\end{center}
\end{figure}

\rm
Applying Lemma~\ref{lemma3} to the set of $(r^2-1)$ possible $\boldalpha\neq\boldgamma$ from\refeq{Galpha_concentration} yields a probabilistic bound on the relative aliasing error when $Z_\boldgamma \neq 0$\footnote{Note that one should first check that every term in the products are positive to ensure that the inequality above be relevant, which will be guaranteed by the final criterion.}:
\begin{multline}
	P\left(
	\left| \sum_{\boldalpha\neq\boldgamma}  \frac{\tilde{Z}_\boldalpha}{\tilde{Z}_\boldgamma}  G_\boldalpha (\k') \right| 
	\leq
	\sqrt{2}\sum_{\boldalpha\neq\boldgamma} \left|\frac{\tilde{Z}_\boldalpha}{\tilde{Z}_\boldgamma}\right| 
	\delta_\boldalpha'
	\right)\\
	\geq 
	\left( 1 - 4e^{- c^2 n_\d} \right)^{r^2-1} \label{bound_alias_H}
\end{multline}
Given some desired relative error $p\in(0,1)$ and lower probability $P_2$, one needs to find whether there exists $c=c_2(p)>0$  such that $\forall  \k'\in{\cal D}_{HR}$
\begin{equation}
\label{ineq_c2p}
		\ds{\sqrt{2}\sum_{\boldalpha\neq\boldgamma}  \left|\frac{\tilde{Z}_\boldalpha}{\tilde{Z}_\boldgamma}\right|  
		\underbrace{\left(c\|\q_\boldalpha\|_1\epsilon +  f(\q_\boldalpha,\epsilon_r)\right)}_{\delta_\boldalpha'}	
		\leq p,}
\end{equation}
A necessary condition appears as
\begin{equation}
\label{def_p0}
	p> p_0(\epsilon,r,Z) = \sqrt{2}\sum_{\boldalpha\neq\boldgamma} \left|\frac{\tilde{Z}_\boldalpha}{\tilde{Z}_\boldgamma}\right|  f(\q_\boldalpha,\epsilon_r)
\end{equation}
Then one can define
\begin{equation}
\label{def_c2}
	c_2(p) = \inf_{\q_\boldgamma} \sup_c \left\{c :(\ref{ineq_c2p}) \mbox{ is true for } \q_\boldgamma \right\}
\end{equation}
If $c_2(p)>0$ is well defined, then there exists a minimum number of images per position $n_\d$ such that
\begin{equation}
\label{critere_nd}
	\left( 1 - 4e^{- c_2(p)^2 n_\d} \right)^{r^2-1} \geq P_2,
\end{equation}
that is
\begin{eqnarray}
\label{nd_min_Hoeffding}
	n_\d^{min}
	& = & \frac{1}{c_2(p)^2}\log \left(\frac{4}{ 1- P_2^{\frac{1}{r^2-1}}}\right)\\
\end{eqnarray}
In the special case $r=2$, \refeq{Galpha_concentration_r2} yields the even tighter bound:
\begin{equation}
\label{nd_min_Hoeffding_r2}
	n_\d^{min} = \frac{1}{c_2(\sqrt{2}p)^2}\log \left(\frac{2}{ 1- P_2^{\frac{1}{3}}}\right).
\end{equation}
Finally, one obtains a probabilistic bound on the aliasing error relative to $|\tilde{Z}(\q_\boldgamma)|$ as 
\begin{equation}
\label{bound_alias_H_gen}
	P\left(
	\left| \sum_{\boldalpha\neq\boldgamma} \frac{\tilde{Z}_\boldalpha}{\tilde{Z}_\boldgamma} G_\boldalpha (\k') \right| 
	\leq
	p
	\right)
	\geq
	P_2
\end{equation}
This relative error provides a good estimate of the relative error on the final restored image under an {\em assumption of perfect deconvolution} in\refeq{X_decomp2} when $\tilde{H}(\k')\neq 0$ everywhere so that $\tilde{Y}(\k') = \tilde{X}(\k')/\tilde{H}(\k')$. 
This result permits to evaluate the contribution of aliasing errors to the reconstructed blurred HR image $Z$. This necessitates knowledge of the true HR image : one can also use the reconstructed image {\em a posteriori} to indicate which frequencies are most suspected to contribute to aliasing effects. Each specific image has a specific structure in the Fourier domain so that special aliasing effects may appear and make superresolution difficult, at least for a small set of frequencies for which the sum of aliasing terms in\refeq{bound_alias_H_gen} may be particularly large.
To propose a generic {\em a priori} estimate of the order of magnitude of this aliasing error, we need to make some assumption on the content of images. It is well accepted that natural images often exhibit a power law energy spectrum $\propto 1/\|\k'\|_2^{2(1+\eta)}$ where usually $|\eta| \ll 1$ \cite{mg01,rb94,c07pami}. Then 
\begin{equation}
\label{Zalpha_sur_Zgamma}
	 \left|\frac{\tilde{Z}_\boldalpha}{\tilde{Z}_\boldgamma}\right| = \frac{|\tilde{H}(\q_\boldalpha)|}{|\tilde{H}(\q_\boldgamma)|}   
	\left(\frac{\|\q_\boldgamma\|_2}{\|\q_\boldalpha\|_2}\right)^{1+\eta} 
\end{equation}
Therefore the strongest constraints appear for high frequencies (large $\k'$ or $\q_\boldgamma$). Note the dependence on the blur kernel which acts as a low-pass filter: the presence of $\tilde{H}$ in\refeq{Zalpha_sur_Zgamma} will have adverse effects. Since we are searching for lower-bounds, forthcoming computations consider the most favourable case when $\tilde{H} = 1$. See section~\ref{numexp} for a numerical illustration of the effect of a realistic Gaussian blur kernel.  
As a consequence, an approximate computation (see Appendix~\ref{computing_c2}) shows that the highest frequencies define $c_2^*(p)$ as 
\begin{equation}
\label{def_c2_PL}
	c_2^*(p) = \frac{p-p_0^*(\epsilon,r)}{a^*(\epsilon,H)}
\end{equation}
where
\begin{eqnarray}
p_0^*(\epsilon,r) & \simeq & b_0\sqrt{2}^\eta\pi^2\epsilon^2 r^2(r^2-1)  \label{p0_explicit}\\ 
a^*(\epsilon,H) & = & \sqrt{2}\sum_{\boldalpha\neq\boldgamma}
	\frac{|\tilde{H}(\q_\boldalpha)|}{|\tilde{H}(\q_\boldgamma)|}   
	\left(\frac{\|\q_\boldgamma\|_2}{\|\q_\boldalpha\|_2}\right)^{1+\eta} 
	\|\q_\boldalpha\|_1\epsilon \nonumber\\
	& \simeq & a_0 2^{1+\eta/2} \epsilon (r^2-1) \quad \mbox {(if } \tilde{H} = 1) \label{a_explicit}
\end{eqnarray}
where the factor $(r^2-1)$ corresponds to the number of aliasing terms; the coefficient $b_0\simeq 2/3$ for $r=2$ and $b_0\simeq 1.2$ for $r\geq 3$ and it is almost independent of the size $N$ of the image for $N\geq 32$; $a_0\simeq 0.63$ for $r=2$ and $a_0\simeq 1.3$ for $r\geq3$ (see Appendix~\ref{computing_c2}). In the general case, \refeq{p0_explicit} \&\refeq{a_explicit} interestingly permit to make explicit the dependence on $r$, $\epsilon$ and $\eta$.
Thus, for a power-law spectrum image, the required minimum number of images/position is:
\begin{eqnarray}
\label{nd_min_Hoeffding_PL}
	n_\d^{min}
	& = & \ds{\frac{a^{*2}(\epsilon,H)}{(p-p_0^*(\epsilon,r))^2}\log \left(\frac{4}{ 1- P_2^{\frac{1}{r^2-1}}}\right)}
\end{eqnarray}
 One observes that $p_0/\epsilon^2r^2$ essentially depends on $r$ as soon as $\epsilon$ is small enough. Figure~\ref{eps_p_bound} illustrates numerical orders of magnitude of reachable $(p,\epsilon)$ such that $p>p_0^*(\epsilon,r)$ for given $r$ under the assumption of a power law spectrum. Pairs of acceptable parameters $(p,\epsilon)$ for which guaranted error bounds exist are at the bottom right of each curve. 
Typical values can be evaluated numerically. For instance assuming $\eta=0$, to guarantee an error smaller than 10\%,  $r=2,p=0.1\Rightarrow \epsilon \leq 0.036$ or $r=6,p=0.1\Rightarrow \epsilon \leq 0.0035$. 
Observe that $\epsilon$ should rapidly decrease as $r$ becomes larger when some given error level $p$ with high probability is desired.
Note the logarithmic dependence on $(1- P_2^{\frac{1}{r^2-1}})$ which permits to choose $P_2$ close to 1 without increasing $n_\d^{min}$ a lot. 

By analyzing our results in the other way, one can also deduce a map of confidence intervals $p(\q)$ for fixed $n_\d$. In practice, one may be constrained by the acquisition protocole, so that $n_\d$ is fixed. Then one can set the value of $c_2(p)$ in\refeq{ineq_c2p} and compute a map of confidence intervals $p(\q)$ in the Fourier domain, taking into account the spectrum of the targeted HR image. Since the true HR image is not known in practice, its Fourier transform may be replaced by its estimate. This procedure helps identifying which frequencies are more likely to contribute to aliasing errors.

\subsection{Main results}
\label{mainresults}

The analysis of the estimate $X$ of the blurred image $Z=HY_{HR}$ by the proposed algorithm gives (see\refeq{X_decomp2}):
\begin{equation}
\label{X_decomp2_theo}
\hspace*{-3mm} \tilde{X}(\k')=\tilde{Z}(\k')G_{\boldgamma}(\k') \\+\underbrace{\sum_{\boldalpha\neq\boldgamma}\tilde{Z}(\k+\boldalpha N) G_{\boldalpha}(\k')}_{B(\k')}
\end{equation}
Theorem~\ref{theorem_alias} below gathers the necessary assumptions on the acquisition system ($r$, $\epsilon$, $\E[\bop_{\d j}$), the scenes (spectrum and $\eta$) and the desired confidence level ($p_1$ \& $P_1$, $p_2$ \& $P_2$) to obtain two fundamental concentration inequalities for the approximation and the aliasing terms respectively.

\begin{theorem}
\label{theorem_alias}
~\\
	{\em Acquisition system - }
	Let $r$ the superresolution factor.
	Let $0<\epsilon<1/\pi r$ the maximum error of the positioning system (in LR pixel units). Assume bounded errors $\bop_{\d j}$ on positions within $(-\epsilon r,\epsilon r)$ in both $x$ and $y$ directions with a possible constant bias $\E[\bop_{\d j}]$ (in HR pixel units). Assume that $n_\d$ images are taken for each one of the $r$ necessary reference positions corresponding to $\d\in(0,r-1)^2$ HR pixel units.\\ 
	{\em Confidence intervals - }
	Let $p_1\in(0,1)$, resp. $p_2\in(0,1)$, the desired maximum relative error on the main approximation term, resp. the sum of aliasing terms, of the reconstructed image ($p_1$ \& $p_2$ will generally be close to 0).\\
	Let $P_1\in(0,1)$ the desired level of confidence in the relative error $p_1$ due to the main approximation term ($P_1$ will be close to 1 so that $1-P_1$ is close to 0).\\
	Let $P_2\in(0,1)$ the desired level of confidence in the relative error $p_2$ due to the aliasing term ($P_2$ will be close to 1).\\
	{\em Technical assumptions - } 
	Assume that one can define $c_1>0$ and $c_2>0$ by (dependences are omitted)
\begin{equation}
\label{c1_th}	
	c_1 = \frac{1}{2\sqrt{2}\pi\epsilon} \left( p_1 - \sqrt{2}\pi\langle\|\E[\bop_{\d,j}]\|_2\rangle_\d - 2\pi^2\epsilon^2 r^2\right) 
\end{equation}
\begin{equation}
	c_2(p_2) = \inf_{\q_\boldgamma} \sup_c \left\{c : L_\boldgamma(c) \leq p_2 \right\} 
\end{equation}
where function $f$ is defined by\refeq{def_f} and
\begin{equation}
	L_\boldgamma(c)  = \sqrt{2}\sum_{\boldalpha\neq\boldgamma}  \left|\frac{\tilde{Z}_\boldalpha}{\tilde{Z}_\boldgamma}\right|  
	\left(  c\|\q_\boldalpha\|_1\epsilon +  f(\q_\boldalpha,\epsilon_r)	\right) \\ 
\end{equation}
If 
\begin{equation} 
\label{bound_approx}
	n_\d \geq \frac{8}{c_1(p_1)^2}\log \left(\frac{4}{1-P_1}\right)
\end{equation}
then the following probabilistic inequality holds:
\begin{equation}
		P\left(\left\{\forall \k'\in{\cal D_{HR}},\: \left|\frac{G_\boldgamma(\k')}{r^2}-1\right| \leq p_1\right\} \right)  \geq  P_1
\end{equation}
If 
\begin{equation}
\label{nd_min_Hoeffding_theo}
	n_\d
	 \geq 
	 \left\{
	 \begin{array}{l}
		 \ds{\frac{1}{c_2(\sqrt{2}p)^2}\log \left(\frac{2}{ 1- P_2^{\frac{1}{3}}}\right) \mbox{ if } r=2,}\\
		 \ds{\frac{1}{c_2(p)^2}\log \left(\frac{4}{ 1- P_2^{\frac{1}{r^2-1}}}\right) \mbox{ if } r\geq 3.}
	 \end{array}
	 \right.
\end{equation}
then the following concentration inequality holds:
\begin{equation}
	P\left(\left\{\forall \k'\in{\cal D_{HR}},\: \left|\frac{B(\k')}{\tilde{Z}(\k')}\right| \leq p_2\right\} \right) 
	\geq  P_2  
\end{equation}


\end{theorem}

\noindent{\em Let us comment on Theorem~\ref{theorem_alias}.}
In ideal experimental conditions, with no positioning bias and $\epsilon r\leq \sqrt{p/2\pi^2}$, 
\begin{equation}
\label{c1_th_approx}
	c_1(p)\simeq \frac{p-2\pi^2\epsilon^2 r^2}{2\sqrt{2}\pi\epsilon}
\end{equation}
The quantity $c_2(p_2)$ can be computed numerically for some given specific image. A necessary condition to the existence of $c_2(p_2)$ is
\begin{equation}
\label{def_p0_theo}
	p_2> p_0(\epsilon,r,Z) = \sqrt{2}\sup_{\q_\boldgamma}\sum_{\boldalpha\neq\boldgamma} \left|\frac{\tilde{Z}_\boldalpha}{\tilde{Z}_\boldgamma}\right|  f(\q_\boldalpha,\epsilon_r)
\end{equation}
In the most favourable case when $\tilde{H}=1$ (no blur) and the image has a power law Fourier spectrum $\propto \|\k'\|_2^{-2(1+\eta)}$,\refeq{p0_explicit} permits to estimate $p_0(\epsilon,r,Z)$.
Then $c_2(p_2)$ can be computed from\refeq{def_c2_PL} which is easy to use and gives quantitative indications about $n_\d$.

\begin{corollary}
\label{corol}
	Under the assumptions of Theorem \ref{theorem_alias} and denoting $c_1=c_1(p_1)$ and $c_2=c_2(p_2)$, if a sufficient number $n_\d$ of images per position is used, one has the following concentration inequality which guarantees a small relative error with high probability:
\begin{multline}
\label{corol_eq}
	P\left(\left\{\forall \k'\in{\cal D_{HR}},\:\left| \frac{\tilde{X}(\k')-\tilde{Z}(\k')}{\tilde{Z}(\k')}\right| \leq 
	p_1+p_2
	\right\} \right)\\
	\geq 
	P_2-(1-P_1)\\
	\geq
	\left(1-4e^{-c_2^2 n_\d}\right)^{r^2-1} - 4e^{-\frac{c_1^2 n_\d}{8}}
\end{multline}
\end{corollary}
{\em Proof :} this is a direct consequence of lemma~\ref{lemma1} applied to the sum of the approximation and aliasing terms.

Corollary~\ref{corol} gives a probabilistic bound to the total relative error on each frequency component of the reconstructed blurred image $Z$ using the algorithm from~\cite{eh01} before the deconvolution step. Note that the bound in probability in\refeq{corol_eq} tends to 1 exponentially fast when $n_\d\to\infty$. In practice, one can guarantee a global relative error $\leq 10\%$ with probability $\geq 0.90$ by choosing $(p_i,P_i)=(0.05,0.95)$, i=1,2.  This result provides a precise quantitative analysis of the reconstruction error. 
One limitation of the present study is that $\tilde{Z}(\k')=H(\k')Y_{HR}(\k')$ is the blurred super-resolved image resulting from the fusion of LR images. However the deconvolution step is common to every acquisition system and remains a limitation of any SR approach. Of course, the most favourable situation is when $H(\k')$ is close to 1, corresponding to a Dirac PSF in the spatial domain. Then Corollary~\ref{corol} gives a good indication of the quality of high resolution imaging by using multiple acquisitions per positions.

In summary, we propose a detailed analysis of the reconstruction error of a fast method in the Fourier domain. It provides an a priori estimate of the number of images/position necessary to guarantee a given quality of reconstruction of each frequency (Fourier mode) with high probability. Based on Monte Carlo simulations, it also allows to estimate a posteriori a map of confidence levels in the frequency domain. Section~\ref{numexp} will show numerically that these bounds are tight. We have worked on the intermediate reconstructed image $Z$ but it appeared that this study of {\em relative} errors produces error estimates for the restored image $Y$ itself under a {\em perfect deconvolution assumption}. Theorem~\ref{theorem_alias} can be used based on the generic assumption of a power-law spectrum that is usual for natural images or more specifically for one specific image.

\subsection{What about the SNR ?} 
\label{control_SNR}

\begin{figure}
\begin{center}
	\includegraphics[width=0.4\textwidth]{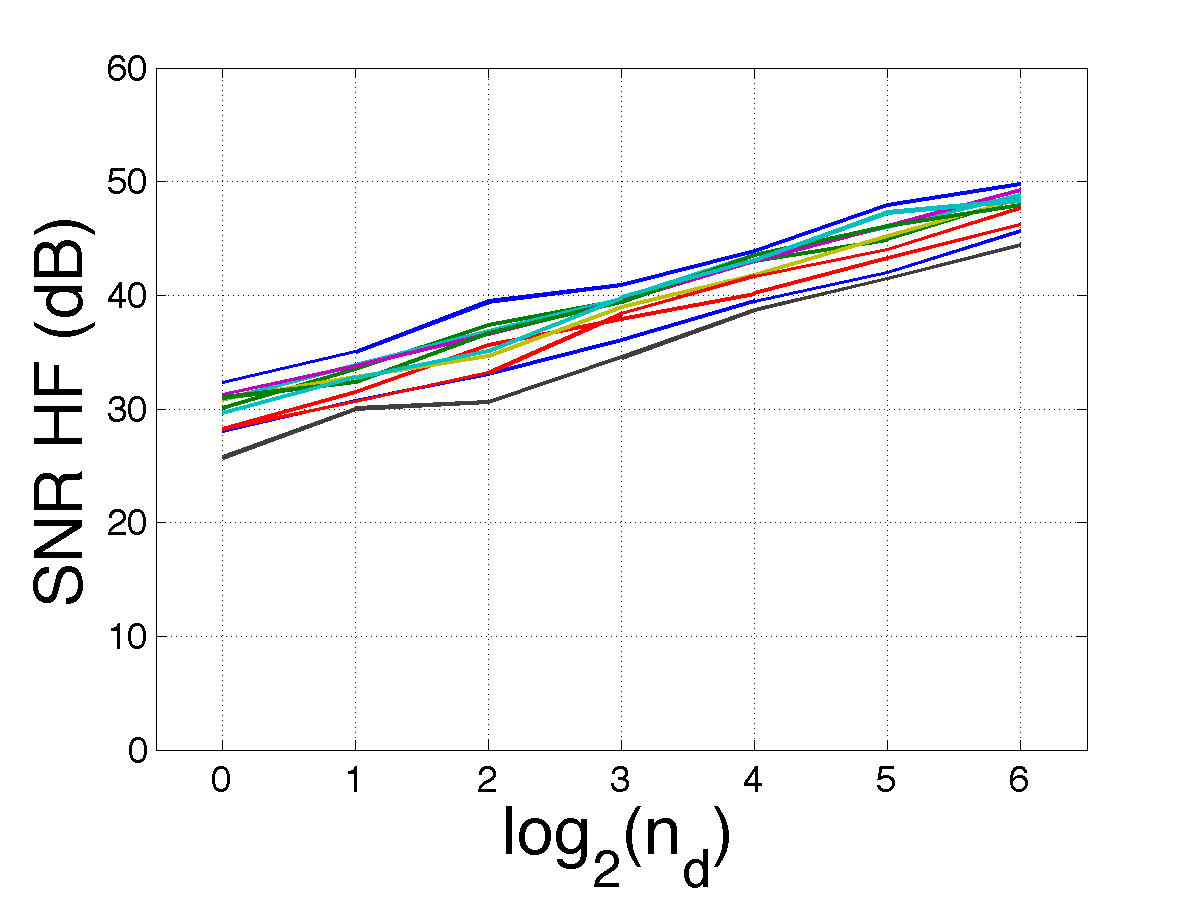}
	\caption{\label{fig_snr_nd} SNR for high frequencies only is proportional to $\log_{10} n_\d$. Results from 100 Monte-Carlo simulations with uniform distribution of positions with $\epsilon=0.01$ for 11 images (Lena, Barbara, Boat...).}
\end{center}
\end{figure}

We have demonstrated theoretical bounds to control the quality of the super resolved image in the Fourier domain. However this result deals with each frequency separately. Now we aim at identifying the dependence of the SNR between the reconstructed image and the ground truth. Again, this SNR deals with $Z$ not $Y_{HR}$ and it measures the quality of the fusion step and does not consider the posterior deconvolution effects.  
We consider the mean square error :
\begin{equation}
\label{MSE}
	\left\| \tilde{X}-\tilde{Z}\right\|_2^2= \sum_k \underbrace{\left| \frac{\tilde{X}(\k')-\tilde{Z}(\k')}{\tilde{Z}(\k')}\right|^2}_{\alpha_{\k'}^2}\cdot |\tilde{Z}(\k')|^2
\end{equation}
and compare it to the energy of the original HR image. The $|\tilde{Z}(\k')|$ are considered as fixed (the ground truth) while the $\alpha_{\k'}$ are random variables here (relative error estimates). Now we show that $\E \alpha_{\k'}^2$ is of the order of $1/n_\d$ for all $\k'$ so that $SNR\propto \log n_\d$.
From\refeq{corol_eq} in Corollary~\ref{corol},
\begin{equation}
	P(|\alpha_{\k'}|\leq p) 
	\geq 1-4r^2e^{-c^2n_\d}	
\end{equation}
where $c^2(p)=\min(c_2^2(p),c_1^2(p)/8)$.
Note from \refeq{def_c2_PL} \&\refeq{c1_th_approx} that the typical order of magnitude of $c_1(p)$ and $c_2(p)$ is $p/\epsilon$ so that we can consider that there exists $\lambda>0$ such that $c^2\geq\lambda p^2/\epsilon^2$. Then 
\begin{eqnarray}
	\E[\alpha_{\k'}^2]
	&\leq& \hspace{-3mm}
	\int_{|\alpha_{\k'}|\leq p} \alpha_{\k'}^2 p(\alpha_{\k'})d\alpha_{\k'}
	  +  \int_{|\alpha_{\k'}|\geq p} \alpha_{\k'}^2 p(\alpha_{\k'})d\alpha_{\k'}\nonumber\\
	&\leq&
	p^2 + 2 \sum_{n=1}^\infty \int_{np}^{(n+1)p} \alpha_{\k'}^2 p(\alpha_{\k'}) d\alpha_{\k'}\nonumber\\
	&\leq&
	p^2 + 4r^2\sum_{n=1}^\infty e^{-\lambda n^2 p^2n_\d/\epsilon^2}(n+1)^2p^2\nonumber\\
	&\leq&
	p^2 (1 + e^{-\lambda p^2n_\d/\epsilon^2} K(n_\d))
\end{eqnarray}
where $K(n_\d)$ is finite, decreasing with $n_\d$ and independent of $\k'$. Choosing $p^2 = 1/n_\d$, one obtains for all $\k'\in{\cal D}_{HR}$,
\begin{equation}
	\E[\alpha_{\k'}^2] \leq \frac{1}{n_\d}(1+e^{-\lambda/\epsilon^2}K(n_\d))
\end{equation}
and consequently taking the expectation of\refeq{MSE},
\begin{multline}
	\E\left[\left\| \tilde{X}-\tilde{Z}\right\|_2^2\right] = \sum_{\k'}  \E[\alpha_{\k'}^2] |\tilde{Z}(\k')|^2 \\
	\leq \frac{1}{n_\d}(1+e^{-\lambda/\epsilon^2}K(n_\d))  \|\tilde{Z}\|_2^2 
\end{multline}
Finally, using Parceval's equality we get:
\begin{equation}
\label{SNR_nd}
	\mbox{SNR}(\tilde{X},\tilde{Z}) \geq 10\log_{10} n_\d + K 
\end{equation}
where $K$ is a constant depending on the energy of the original image. 
As far as the influence of the number of images per position $n_\d$ is concerned, one can keep in mind that the SNR is improved with a magnitude of 10dB/decade. We can compare this result with the Cramer-Rao lower bound $T_{weak}\propto 1/\sqrt{K+1}$ where $K+1$ is the number of images in \cite{rm06a} : at best, the SNR can grow as log(number of images) which is what\refeq{SNR_nd} predicts. This indicates that the proposed method is efficient at the best expected level \cite{nb02,nb03, bkz94,bk02,tla12,cbk09,rm06a,th84}. Fig.~\ref{fig_snr_nd} shows SNR computed for high frequencies only (the reconstructed HR part of the spectrum). Results were computed from 100 Monte-Carlo simulations with uniform distribution of positions with $\epsilon=0.01$ for 11 images (Lena, Barbara, Boat...). This result completes previous ones in terms of probabilistic bounds on each term of the Fourier transform which are much stronger since they guarantee that {\em all} Fourier modes are accurately recovered with high probability. These detailed bounds permit to prove a global lower bound on the SNR $\propto \log n_\d$. 

\section{Numerical results}
\label{numexp}

To illuminate the complex interplay between the many parameters involved, we study the problem from various viewpoints. 
Section~\ref{illustre_theorem} studies the lower-bound on the number $n_\d$ of images per position to guarantee a given maximum errror level. Section~\ref{tightbounds} compares our theoretical results to numerical estimates of probabilities from Monte-Carlo simulations. Section~\ref{spatial} studies the connection between results in the Fourier domain and their practical impact in the spatial domain. Section~\ref{noise_PSF} shows how the presence of noise and the nature of the blur operator influence the results. Monte-Carlo simulations use 100 realizations of the acquisition procedure assuming a uniform distribution of position errors in $(-\epsilon,\epsilon)$. When no image is specified, the power law spectrum assumption is used.

\subsection{How many images to guarantee some given maximum error level ?}
\label{illustre_theorem}

\begin{figure}
\begin{center}
\includegraphics[height=40mm]{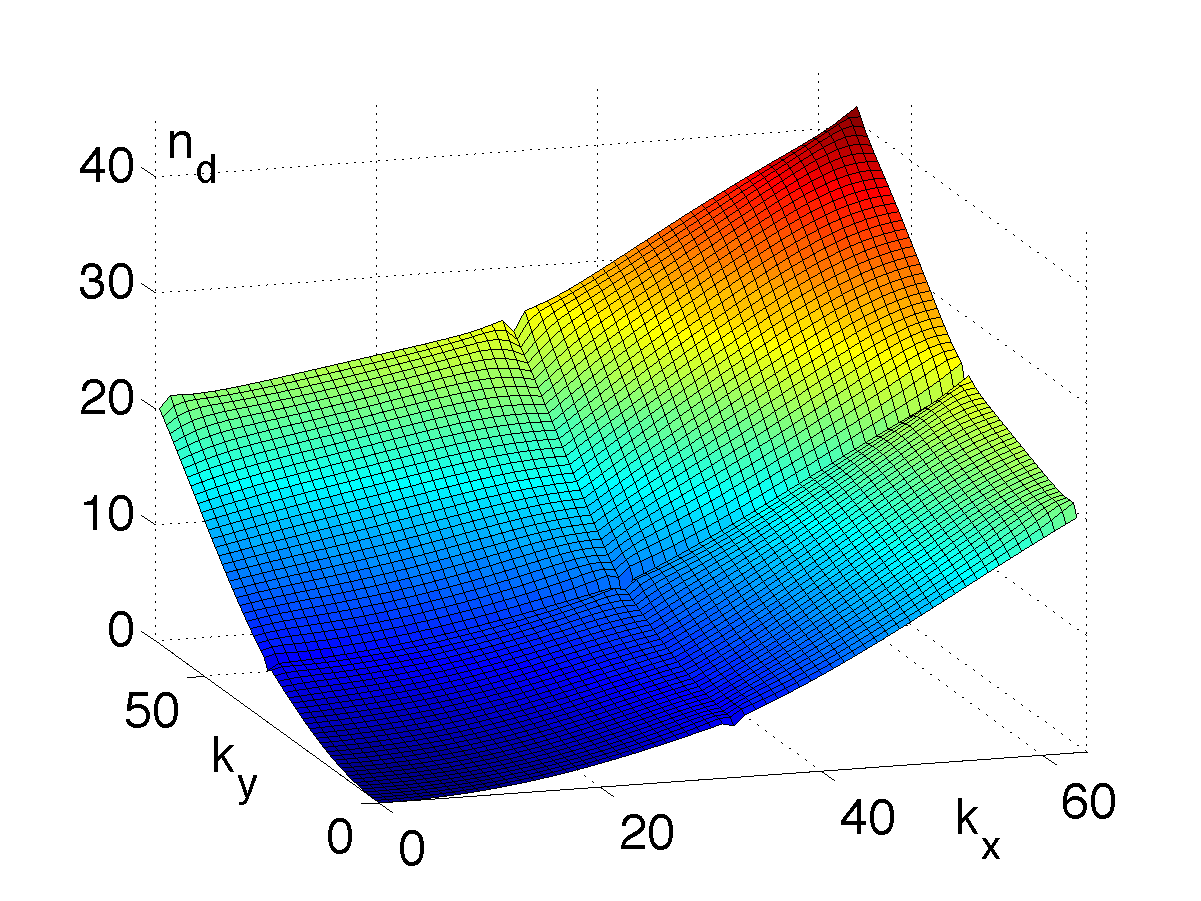}
\caption{\label{fig_nd_k} Minimum number $n_\d(\k')$ to guarantee an aliasing error $\leq 5\%$ with probability $\geq 0.95$ for all $\k'$ ; $\epsilon=0.001$, $r=4$.}
\end{center}
\end{figure}
Fig.~\ref{fig_nd_k} illustrates  the dependence of the required number of images $n_\d(\k')$ on the frequency $\k'$ to guarantee that aliasing contribution is less than 5\% with probability $\geq 0.95$ when $r=4$ and $\epsilon=0.001$ for an image with a power law spectrum. As expected, the recovery of high frequencies requires more LR images. The results are nearly independent of the size $N$ of images as soon as $N\geq 32$. A similar picture (not shown) stands for the main approximation term. 
In general though not always, the control of aliasing effects is the most constraining.

Tab.~\ref{table1} gathers the constraints for various values of $r$ and $\epsilon$ for images with a power law spectrum ($\eta=0$ here). Numbers are computed from\refeq{bound_approx} \&\refeq{nd_min_Hoeffding_theo} in Theorem~\ref{theorem_alias} for parameters $(p_i,P_i)$ = (0.05,0.95), $i=1,2$. This choice of equidistribution of error is certainly not optimal but of practical use with respect to Corollay~\ref{corol} which then permits to guarantee an error level $\leq p_1+p_2=0.10$ with probability larger than $P_2-(1-P_1)=0.90$. 
The larger $r$, the larger the need for multiple images. The smaller the positioning uncertainty $\epsilon$, the smaller the lower bound on $n_\d$. 
In our microscopy setting, 1 LR pixel $\simeq$ 100~nm. The random bias on the platform positioning system is between $0.1$ and 1~nm that is $\epsilon\simeq 0.001-0.01$ LR pixel. The acquition of 1 image usually takes between 0.1 and 0.5~s. In practice, $r^2$ displacements are used so that a minimum acquisition time of about  $r^2\times n_\d\times 0.1$s is necessary.
For instance, for $r=2$ and considering aliasing terms only,  when $\epsilon=0.001$, resp. $\epsilon=0.01$, an aliasing error $\leq 5$\% on the restored image can be guaranteed with probability $\geq 0.95$ by using at least $n_\d=1$, resp. $n_\d=64$, images/position. 
Taking into account both contributions (approximation + aliasing) for $r=2$ and $\epsilon=0.01$, one gets that $n_\d\geq \max(157,64)$ so that $n_\d=157$ images/position are necessary. At 0.1s/im, the acquisition time is of about $r^2\times 157\times 0.1= 64$s. 
If the position accuracy is not better than $\epsilon=0.01$, the potential for superresolution is very limited. Yet for $r=3$ there is no way to guarantee a quality of reconstruction with an aliasing error less than $5\%$ (NR = "Not Reachable" in Tab.~\ref{table1}). However, for sufficiently accurate positioning $\epsilon=0.001$ the acquisition of $r^2\times 13 = 117$ images ($\simeq$ 59s at 0.5~s/im. or 12s at 0.1s/im.) permits to guarantee a relative error $\leq 10\%$ with probability $\geq 0.90$ for all frequencies. 
For $r=4$, more than 42 images/position are necessary which leads to an acquisition time of about $r^2\times 42\times 0.1\simeq 67$s that is still reasonable in many contexts.
For $r=6$, $r^2\times 42= 18324$ images would be necessary ($\simeq$ 30 min at 0.1~s/im.) which becomes technically difficult, even regardless of other physical limitations of the system which make the objective $r=6$ unrealistic.
Remember that these predictions on $n_\d$ are based on the generic assumption of a power law spectrum which is statistically common to many natural images. In full rigor, even though these numbers are of great use in practice to calibrate the acquisition system, they should be estimated for each image individually: then the full map of the bounds in the Fourier domain can be computed. 

\begin{table}
\begin{center}
{
\setlength{\tabcolsep}{2mm}
\begin{tabular}{|c|r@{~/~}r|r@{~/~}r|r@{~/~}r|}
\hline\hline   
\multicolumn{7}{|c|}{images with a power law spectrum}\\
\hline\hline   
 $\epsilon$   &  \multicolumn{2}{c|}{0.01}          & \multicolumn{2}{c|}{0.001}     & \multicolumn{2}{c|}{0.0001}\\
		 &  approx. & alias	&  app. & alias	&  app. & alias\\	
 \hline
 $r= 2$     &   157  &  64 	 & 2  &  1         & 1  &  1\\
 PSF(0.5) &  157 &  6108 	& 2  &  23      & 1   &  1\\
 \hline
 $r=3$         & 267  &  NR    &    2  &  13      &  1  &   1    \\
 PSF(0.5) &  267  &  NR 	& 2  &  540       &1    &  5\\
 \hline
 $r= 4$     &  817  &  NR       &  2  &  43       & 1  &  1 \\
 PSF(0.5) & 817  &  NR 	& 2  &  2228        & 1   &  13\\
 \hline
 $r=5$ & 651162  &   NR     &    2  &   168     &   1  &  2   \\
 PSF(0.5) &  651162 &  NR 	& 2  &  73025   & 1   &  43\\
 \hline
 $r= 6$  	&  NR  &  NR       &  2  &  516       & 1  &  3\\
 PSF(0.5) &   &  NR 	& 2  &  NR   & 1   &  82\\
 \hline
 $r=7$      & NR  &  NR       &   2  &   3486    &   1  &  6  \\
 PSF(0.5) &  &  NR     & 2  &  NR         & 1   &  189\\
 \hline
 $r= 8$        &  NR  &  NR   &  2  &  NR         & 1  &  9\\
 PSF(0.5) &  &  NR 	& 2  &  NR      & 1   &  311\\
 \hline\hline
\end{tabular}

}
\caption{\label{table1} Minimum $n_\d$ for accuracy $<5\%$ 
with probability $>0.95$ on approx./aliasing as a function of $\epsilon$ in LR pixel units. PSF(0.5) is Gaussian blur of width 0.5; NR=`Not Reachable'.} 
\end{center}
\end{table}

 
 

\subsection{How realistic and tight are these bounds ?} 
\label{tightbounds}
\begin{figure}
\begin{center}
	\includegraphics[width=0.35\textwidth]{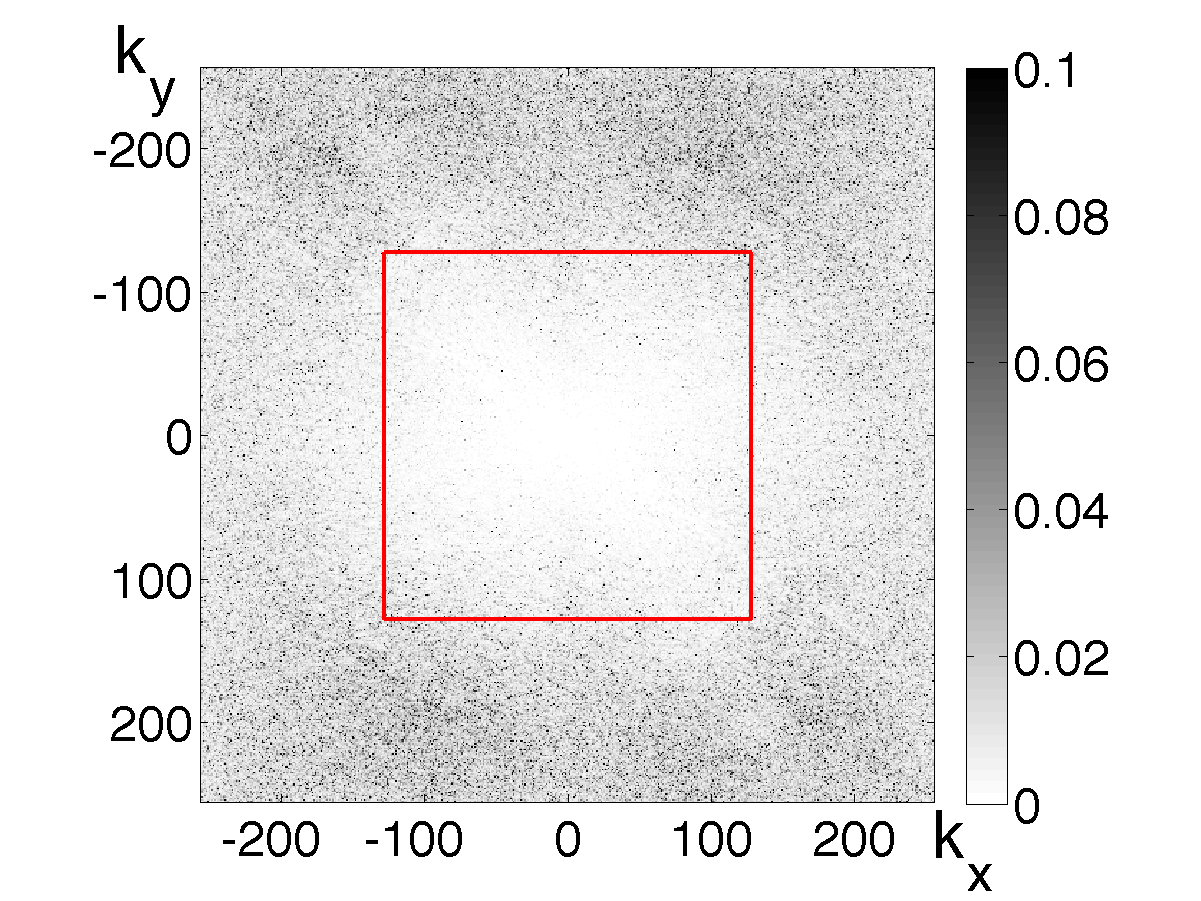}
	\caption{\label{map_p_lena} Fourier map of the lower bound on the aliasing error $p_2$ within $(0,0.1)$ for Lena. The red square indicates LR frequencies. $r=2$ \& $\epsilon=0.01$:  for black points (0.9\% points are $>0.1$), superresolution with guaranted error bound is not feasible (100 MC simulations).}
\end{center}
\end{figure}
\begin{figure}
\begin{center}
	\includegraphics[width=0.23\textwidth]{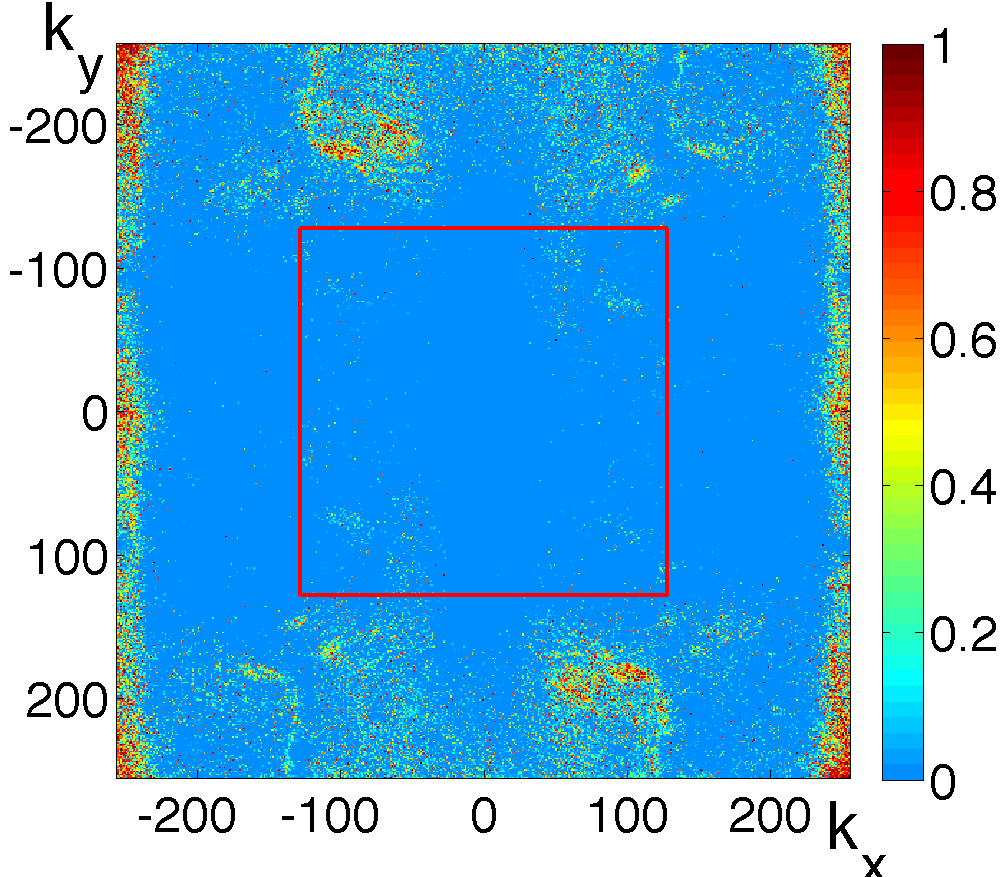}
	\includegraphics[width=0.24\textwidth]{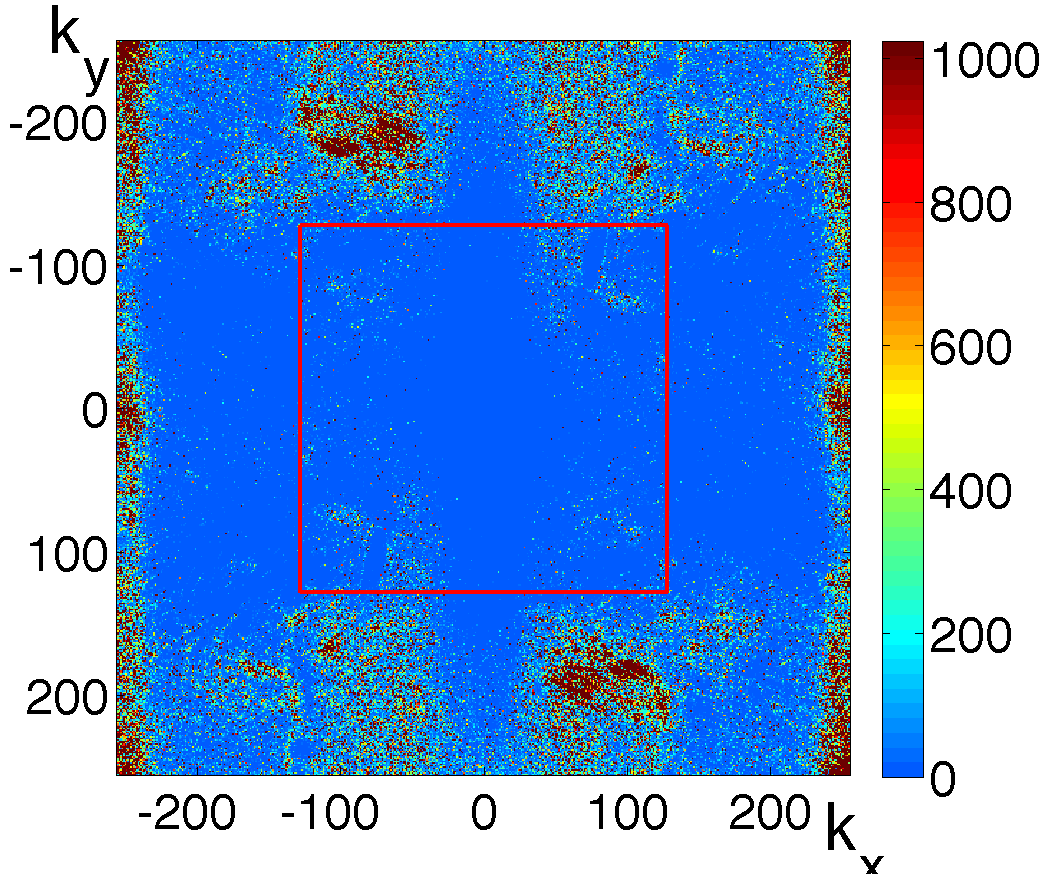}
	\caption{\label{maps_bouquetin} In Fourier domain, for image Barbara, $r=2$ and $\epsilon=0.01$: (left) probability that relative error due to aliasing be $\geq 0.1$ using $n_\d=64$ im/pos. ;  (right) minimum number of images to ensure an aliasing error $<10\%$.}
\end{center}
\end{figure}

When considering one specific image, the bounds from Theorem~\ref{theorem_alias} can first be considered to dimension the acquisition system and then to check the reliability or accuracy of the restored image. As a first step using\refeq{def_p0_theo} we can compute a map of the lower bound on the aliasing error $p_2$ in the HR Fourier domain given the motion accuracy $\epsilon$: this map tells us what is the best achievable relative accuracy for each frequency $\k'$.  
For Lena, $r=2$ and $\epsilon=0.01$, Fig.~\ref{map_p_lena} shows the map computed from 100 Monte-Carlo simulations over uniformly distributed positioning errors in $(-\epsilon,\epsilon)$. Gray points are such that a sufficient number of images/position should permit to guarantee a relative error $<10\%$ with high probability. Few black points where the lower bound of $p_2$ is $>0.1$ correspond to spatial frequencies for which an error $<10\%$ cannot be guaranteed, whatever the number $n_\d$ of images per position mainly because of excessive aliasing. As expected, we observe that high frequencies are the most difficult to reconstruct accurately.
Fig.~\ref{maps_bouquetin}(left) shows in Fourier domain the probability that the aliasing error at $\k'$ be $\geq 10\%$ when using 32 im./position for Barbara with $\epsilon=0.01$ and $r=2$. Fig.~\ref{maps_bouquetin}(right) shows the number $n_\d$ of images necessary to ensure that aliasing error $\leq 10\%$ according to Theorem~\ref{theorem_alias}. Note the consistency between these pictures. Aliasing effects are a direct consequence of the  image spectrum: some frequencies are much more difficult to reconstruct and call for a larger $n_\d$.


\subsection{How are the errors localized in the spatial domain ?}
\label{spatial}

\begin{figure}
\begin{center}
	\includegraphics[width=41mm, height=41mm]{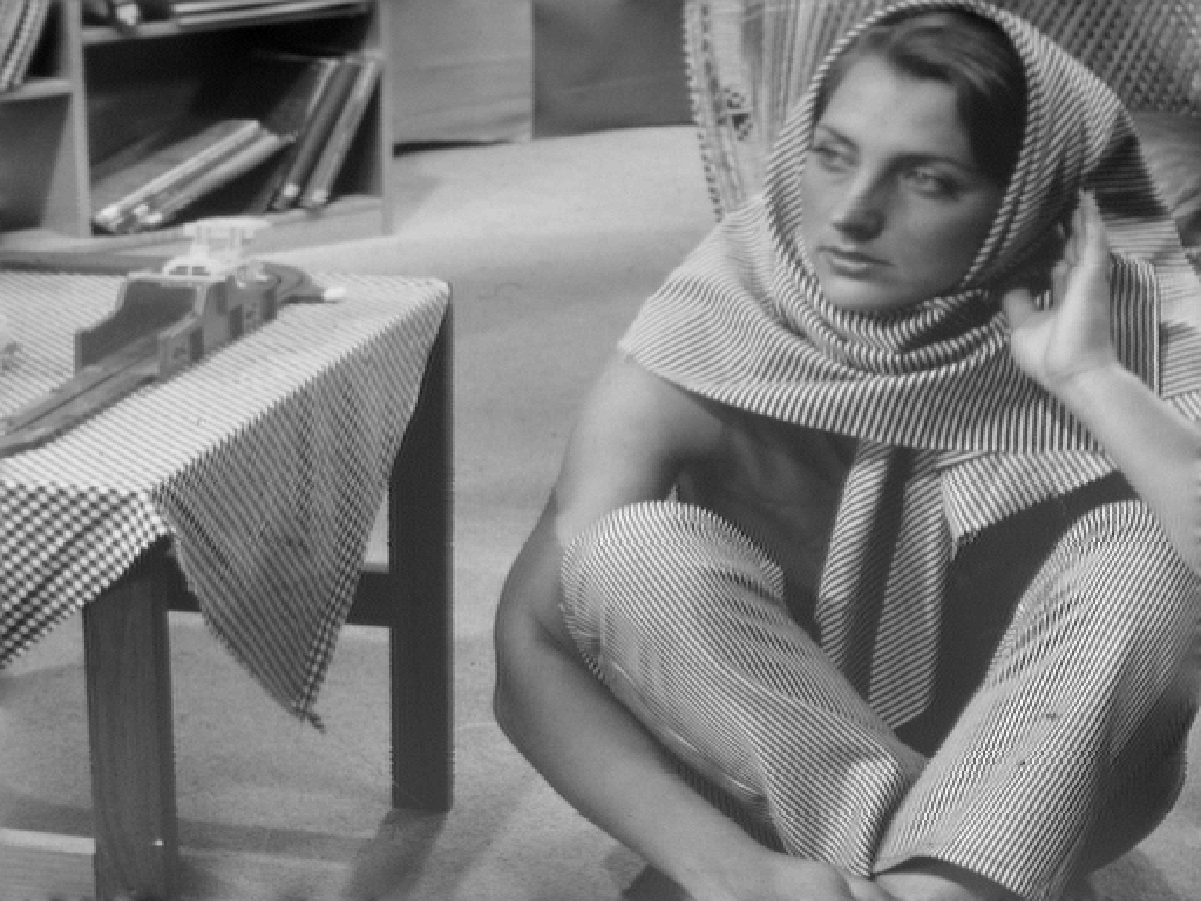}
	\includegraphics[width=41mm, height=41mm]{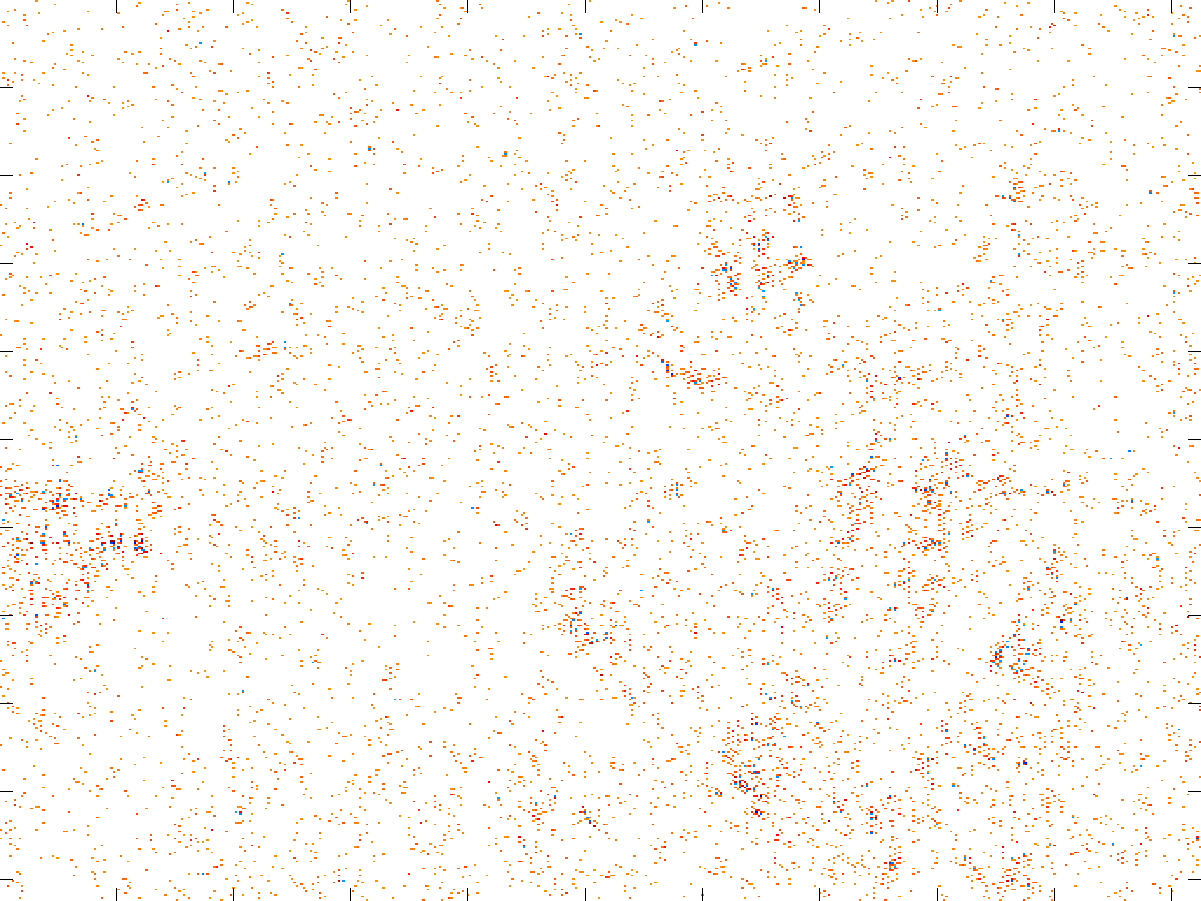}
\caption{\label{aliasing_barbara_spatial} (l.) Barbara, (r.) contribution of the less reliable frequencies.} 
\end{center}
\end{figure}

In the previous sections, we have analyzed errors in the frequency domain. We are now interested in sudying the localization of these errors in the spatial domain. 
 We use Monte-Carlo simulations to estimate the most affected regions. By selecting the less reliable frequency components of an image where the aliasing error is $\geq 10\%$ with probability $\geq 0.1$, one can reconstruct the corresponding spatial counterpart and then localize and quantify their contribution. For $r=2$, $\epsilon=0.01$, the contribution of aliasing to high superresolved frequencies weights for a SNR of -27.2dB. 
The present theoretical analysis permits such a selection of frequencies as well. For a given number $n_\d$ of images/position (which may be insufficient to reliably reconstruct some frequencies) one can reconstruct the spatial counterpart of the less reliable frequencies where the aliasing error is expected to be $\geq 10\%$ with probability $\geq 0.1$ according to Theorem~\ref{theorem_alias}.
Fig.~\ref{aliasing_barbara_spatial} shows such a picture for {\em Barbara} for $r=2$, $\epsilon=0.01$ and $n_\d=256$ im./pos., to be compared with the minimum $n_\d=157$ in Tab.~\ref{table1}. As expected, the spoiled regions are the most textured ones as well as some contours (better seen on screen). Remember that the analysis focused on the modulus of Fourier spectra while phases carry the location information. Maximum amplitudes are of about 4 and the standard deviation is of 0.67 (to compare with 255 in 8 bits). These "non reliable" components then weights for a SNR of -25.8~dB w.r.t. superresolved frequencies only. 
At least on this example, our theoretical predictions both qualitatively and quantitatively agree with Monte Carlo results.
We emphasize that our analysis not only gives indications to chose $n_\d$ but also produces a detailed map of  errors distribution both in the Fourier domain and in the spatial domain. 


\subsection{How do noise and PSF influence performances ?}
\label{noise_PSF}

Two important questions remain: how does noise would impact the present approach ? how does the PSF influence results ?
The problem of noise is not the most critical: averaging numerous images attenuates additive noise. The present approach considers additive contributions of numerous images affected by independent realizations of noise: this naturally tends to increase the signal to noise ratio. This is easily checked experimentally and not illustrated here for sake of briefness. 
The question of the PSF is a much bigger concern since it is involved in the error analysis until the end. Of course frequencies where $\tilde{H}(\k')=0$ are lost (except using some a priori during the restoration step) and we already mentionned that the present analysis is no more valid for these frequencies. Moreover the structure of aliasing is influenced by the PSF in an important manner, see\refeq{Zalpha_sur_Zgamma}. All the experiments above considered the ideal situation of a Dirac PSF where $\tilde{H}(\k')=1$ $\forall\k'$. The lines `PSF(0.5)' in Tab.~\ref{table1} show how the lower bounds of $n_\d$ are modified in presence of a Gaussian blur of width 0.5. As expected it dramatically influences the estimates, e.g. for $(r,\epsilon)=(2,0.001)$ the bound becomes 18 in place of 1. It is welknown that the control of the PSF is a real stake in the conception of a superresolution system: the present study permits to quantitatively evaluate its influence.

\section{Conclusion}
\label{conclusion}

We propose a controlled cheap and fast super-resolution technique which takes benefit from (1) the piezoelectric positioning systems of microscopes (or telescopes) to realize accurate translations, (2) the statistical analysis of the fast algorithm proposed in~\cite{eh01} so that error confidence intervals can be computed as a function of the number of available images. This is made possible by the simplicity of the algorithm itself and by exploiting the averaging effect of LR images taken at positions that are randomly distributed around the same reference position. This technique is cheap and realistic to enhance the resolution of many devices. It may be adapted to many scientific applications ranging from biology to astronomy where the need for guarantees on the restored information is crucial. This analysis deals with clean images without noise. Note however that any zero-mean noise gets attenuated in the HR image reconstruction by using more and more images per position. The resulting probabilistic upper bounds are a good complement to the Cramer-Rao lower bounds in~\cite{rm06a} and close to tight since the order of magnitudes are similar. Numerical experiments illustrate our results in both the Fourier and spatial domains as well as the effect of the PSF. One limitation of this work is the choice of a simple setting~\cite{eh01}. Future works should investigate how these results could be extended to more sophisticated superresolution algorithms where different reconstruction priors are used, e.g. \cite{frem04,h07a,vsvv07}. One strong aspect of this work is in its predictions for practical implementation. Applications in microscopy for biological imaging as well as in photo cameras are the subject of ongoing work.

%

%

%

%




\section{Appendix}

\subsection{Proofs of Lemma 1 \& 2}
\label{prooflemmas}

%
%

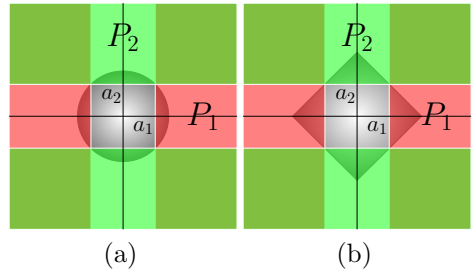
\begin{figure}
\begin{center}
{\setlength{\tabcolsep}{0pt}
\begin{tabular}{ccc}
\scalebox{0.6}{
\begin{tikzpicture}
	\shadedraw[inner color=white,outer color=gray,draw=gray] (0,0) circle (1cm);
	\filldraw[color=white, fill=red, fill opacity=0.5] 	(0.705,-2.5)	rectangle (2.5,2.5);
	\filldraw[color=white, fill=red, fill opacity=0.5] 	(-0.705,-2.5)	rectangle (-2.5,2.5);
	\filldraw[color=white, fill=green, fill opacity=0.5]  (-2.5,0.705)	rectangle (2.5,2.5);
	\filldraw[color=white, fill=green, fill opacity=0.5]  (-2.5,-0.705)	rectangle (2.5,-2.5);
	\node	at (0,1.75) {\huge $P_2$};
	\node	at (1.75,0) {\huge $P_1$};
	\draw (-2.5,0) -- (2.5,0);
	\draw (0,-2.5) -- (0,2.5); 
	\node	at (0.45,-0.25) {\Large $a_1$};
	\node	at (-0.25,0.45) {\Large $a_2$};
\end{tikzpicture}}
&  \hspace{0mm} &
\scalebox{0.6}{
\begin{tikzpicture}
	\shadedraw[inner color=white,outer color=gray,draw=gray, rotate=-45] (-1,-1) rectangle (1,1);
	\filldraw[color=white, fill=red, fill opacity=0.5] 	(0.705,-2.5)	rectangle (2.5,2.5);
	\filldraw[color=white, fill=red, fill opacity=0.5] 	(-0.705,-2.5)	rectangle (-2.5,2.5);
	\filldraw[color=white, fill=green, fill opacity=0.5]  (-2.5,0.705)	rectangle (2.5,2.5);
	\filldraw[color=white, fill=green, fill opacity=0.5]  (-2.5,-0.705)	rectangle (2.5,-2.5);
	\node	at (0,1.75) {\huge $P_2$};
	\node	at (1.75,0) {\huge $P_1$};
	\draw (-2.5,0) -- (2.5,0);
	\draw (0,-2.5) -- (0,2.5); 
	\node	at (0.45,-0.25) {\Large $a_1$};
	\node	at (-0.25,0.45) {\Large $a_2$};
\end{tikzpicture}}\\
(a) &  \hspace{0mm} & (b)
\end{tabular}}
\caption{\label{fig_lemma1} Illustrations for the proof of Lemma \ref{lemma1}.}
\end{center}
\end{figure}

\noindent{\em Proof of Lemma~\ref{lemma1}:} 
$\sqrt{x_1^2+x_2^2}\geq \sqrt{a_1^2+a_2^2}$ $\Rightarrow$ $|x_1|\geq a_1$ or $|x_2|\geq a_2$ proves\refeq{sum_square}, see figure~\ref{fig_lemma1}(a).
$|x_1|+|x_2| \geq a_1+a_2$ $\Rightarrow$ $|x_1|\geq a_1$ or $|x_2|\geq a_2$ proves\refeq{sum_module}, see figure~\ref{fig_lemma1}(b) where the grey lozenge represents the region where $|x_1|+|x_2|\leq a_1+a_2$.
 \hfill {\em QED.}

\noindent{\em Proof of Lemma~\ref{lemma3}:}  $\forall i,\: |x_i|\leq a_i \Rightarrow \sum_i |x_i| \leq \sum_i a_i$ 
so that $P\left(\sum_i |x_i| \leq \sum a_i \right) \geq P\left(\left\{\forall i,\: |x_i|\leq a_i \right\} \right)$. 
Since the $x_i$ are independent, 
$P\left(\left\{\forall i,\: |x_i|\leq a_i \right\} \right) = \prod_i P(|x_i|\leq a_i)$.
Noting that 
$\forall i,\: P(|x_i|\leq a_i)\geq (1-P_i)$ concludes the proof. \hfill {\em QED.}

\subsection{Computing $c_2(p)$ in\refeq{def_c2}}
\label{computing_c2}

Here we estimate an order of magnitude of $c_2(p)$ in\refeq{def_c2} under assumptions of Theorem~\ref{theorem_alias}. 
If one neglects the effect of blur in aliasing, we aim at computing the maximum value of $c_2(p)$ such that for all $\q_\boldgamma$,
\begin{equation}
\label{erreur_p_general_app}
	a \:c_2(p) 
	+ 
	p_0	
	\leq p.
\end{equation}
after little reorganization of\refeq{ineq_c2p} where we use
\begin{eqnarray}
	p_0 & \simeq & \frac{\sqrt{2}}{2}\sum_{\boldalpha\neq\boldgamma} \frac{|Y(\q_\boldalpha)|}{|Y(\q_\boldgamma)|} 
	\|\q_\boldalpha\|_1^2\epsilon_r^2 \label{def_p0_app} \\
	a & = &  \sqrt{2}\sum_{\boldalpha\neq\boldgamma} \frac{|Y(\q_\boldalpha)|}{|Y(\q_\boldgamma)|} 
	 \|\q_\boldalpha\|_1\epsilon.\label{def_a_app}
\end{eqnarray}
as soon as $\epsilon_r\ll 1$ so that cubic terms can be neglected.
We first study\refeq{def_p0_app} to get an estimate of $p_0$ for practical use. Note that we will focus on the highest frequencies only, that is typically $\q_\boldgamma = (\pi-\frac{2\pi}{rN},\pi-\frac{2\pi}{rN})$. As a consequence, note that $\|\q_\boldgamma\|_2^{1+\eta}\simeq (\sqrt{2}\pi)^{1+\eta}$. Then, one needs to detail:
\begin{eqnarray}
	\sum_{\boldalpha\neq \boldgamma} \frac{\|\q_\boldalpha\|_1^2}{\|\q_\boldalpha\|_2^{1+\eta}} 
	& = & 
	\frac{\pi^2}{\pi^{1+\eta}} \underbrace{\sum_{\boldbeta\neq (0,0)} \frac{\|\v_{rN}-2\boldbeta/r\|_1^2}{\|\v_{rN} -2\boldbeta/r\|_2^{1+\eta}}}_{F(r,N)}
\end{eqnarray}
where $\v_{rN} = (1-2/rN,1-2/rN)$. The sum $F(r,N)$ can be computed numerically. It weakly depends on $N$ for $N\geq 32$ so that 
\begin{equation}
F(r,N) \simeq 
\left\{
\begin{array}{ll}
	\ds{2=\frac{2}{3}(r^2-1)} & \mbox{ if } r=2,\\[2mm]
	1.2(r^2-1) & \mbox{ if } r\geq 3.
\end{array}
\right.
\end{equation}
For $r=2$, computations can be made by hand easily so that only 2 terms both equal to 1 appear in $F(r,N)$. For $r\geq 3$, it would become more technical. However, one can observe that $\|\q_\boldalpha\|_1 \sim \|\q_\boldalpha\|_2$ (norms are equivalent) so that when $\eta=0$ one expects that $F(r,N) \propto (r^2-1)$, the number of terms in $\sum_{\boldbeta\neq (0,0)}$. This is due to the fact that $\langle \|\q_\boldalpha\|_1\rangle_{\boldalpha\neq\boldgamma}\simeq \pi$ for large $r$. As a result, one obtains in good approximation that :
\begin{equation}
	p_0 \simeq 
		b_0\sqrt{2}^\eta \pi^2\epsilon^2 r^2(r^2-1) 
\end{equation}
where $b_0=2/3$ if $r=2$ or $b_0\simeq 1.2$ if $r\geq 3$.
Now let study coefficient \refeq{def_a_app} along the same lines.
\begin{equation}
 	 a 
	 \simeq
	 \sqrt{2} \sum_{\boldalpha\neq \boldgamma} \frac{\|\q_\boldgamma\|_2^{1+\eta}}{\|\q_\boldalpha\|_2^{1+\eta}} \|\q_\boldalpha\|_1\epsilon
\end{equation}
Using that $\|\q_\boldalpha\|_1 \sim \|\q_\boldalpha\|_2$ (within constant factors), one expects that when $\eta=0$, 
\begin{eqnarray}
	a & \propto & (\sqrt{2}\pi)^{1+\eta} \sqrt{2}\pi^{-\eta}(r^2-1)\epsilon\\
	& \propto & 2^{1+\eta/2}(r^2-1)\epsilon
\end{eqnarray}
Numerical estimates for various values of $r$ ranging from 2 to 8 show that
\begin{equation}
	a =  a_0 \times 2^{1+\eta/2}\epsilon(r^2-1)
\end{equation}
where $a_0$ varies with $\eta$ around a typical value of 1.3 for $\eta=0$, e.g. $a_0\simeq 0.95$ if $\eta=-0.2$ and $a_0\simeq 1.85$ if $\eta=0.2$ for all $r\geq 3$. For $r=2$, one finds $a_0 \simeq 0.63$, resp. $1.14$ and $3.04$ when $\eta=-0.2$, resp. $0$ and $0.2$.

%

\subsection{Properties of complex roots of unity}
\label{rootsofunity}
We recall some usual properties of complex roots of unity. Let
\begin{equation}
	\theta_{\boldalpha \d} = \frac{2\pi}{r}(\boldalpha-\boldgamma)\d = \frac{2\pi}{r}\bolddelta\d
\end{equation}
where $\boldalpha$ and $\boldgamma$ are pairs of integers in $(0,r-1)^2$. 
The set of the $\theta_{\boldalpha\d}$ matches the set of products of complex roots of unity.
As a consequence one has:
\begin{eqnarray}
\label{sumzero_app}
	\sum_\d \cos\left(\frac{2\pi}{r}(\boldalpha-\boldgamma)\d\right) = \sum_\d \cos(\theta_{\boldalpha \d}) = 0 \label{sumzerocos_app}\\
	\sum_\d \sin\left(\frac{2\pi}{r}(\boldalpha-\boldgamma)\d\right) = \sum_\d \sin(\theta_{\boldalpha \d}) = 0 \label{sumzerosin_app}\\
	\sum_\d \cos^2(\theta_{\boldalpha \d}) = 
	\left\{
	\begin{array}{l} 
		r^2 \mbox{ if }  \ds{\boldalpha-\boldgamma\in \{ 0,r/2\}^2,}\\
		\ds{\frac{r^2}{2}} \mbox{ otherwise}.
	\end{array}
	\right. 
	\label{sumcos2_app}\\
	\sum_\d \sin^2(\theta_{\boldalpha \d}) = 
	\left\{
	\begin{array}{l} 
		0 \mbox{ if }   \ds{\boldalpha-\boldgamma\in \{ 0,r/2\}^2,}\\
		\ds{\frac{r^2}{2}}  \mbox{ otherwise}.
	\end{array}
	\right.
	 \label{sumsin2_app}	
\end{eqnarray}
Properties\refeq{sumzerocos_app} and\refeq{sumzerosin_app} come from the observation that 
\begin{multline}
\label{product}
	\sum_{\d\in(0,r-1)^2} e^{i\theta_{\boldalpha \d}} =\\
	\left( \sum_{d_1\in(0,r-1)} e^{i 2\pi(\alpha_1-\gamma_1) d_1/r} \right)  \left( \sum_{d_2\in(0,r-1)} e^{i 2\pi (\alpha_2-\gamma_2) d_2/r}	 \right) 
\end{multline}
where each factor in the r.h.s. is zero since $\boldalpha\neq\boldgamma$ and for any integer $1\leq \delta \leq r-1$, 
\begin{equation}
\label{usual_prop_root}
	\sum_{d=0}^{r-1} e^{i 2\pi \delta d/r} = \frac{1-e^{i 2\pi\delta}}{1-e^{i 2\pi\delta/r}} = 0 
\end{equation}
Now we prove\refeq{sumcos2_app} and\refeq{sumsin2_app}. To this aim we need:
\begin{eqnarray}
	\cos^2(\theta_{\boldalpha \d}) & = & \frac{1+\cos(2\theta_{\boldalpha \d})}{2} \label{coscarre}\\
	\sin^2(\theta_{\boldalpha \d}) & = & \frac{1-\cos(2\theta_{\boldalpha \d})}{2} \label{sincarre}
\end{eqnarray}
We need to evaluate $\sum_{d=0}^{r-1} e^{i 2\theta_{\boldalpha \d}}$.
For $0\leq \delta \leq r-1$, 
\begin{equation}
\label{usual_prop_root2}
	\sum_{d=0}^{r-1} e^{i 4\pi \delta d/r} = 
	\left\{
	\begin{array}{ll}
		\sum_{d=0}^{r-1} 1 = r & \mbox{ if } \delta\in\{0, r/2\},\\  
		\ds{\frac{1-e^{i 4\pi\delta}}{1-e^{i 4\pi\delta/r}} = 0 }& \mbox{ otherwise, } 
	\end{array}
	\right.	
\end{equation}
so that using\refeq{product} again
\begin{equation}
	\sum_\d e^{i 2\theta_{\boldalpha \d} } =
	\left\{
	\begin{array}{l} 
		r^2 \mbox{ if } \ds{\boldalpha-\boldgamma\in \{ 0,r/2\}^2,} \\ 
		0  \mbox{ otherwise}.
	\end{array}
	\right.	
\end{equation}
Taking the real part yields $\sum_\d \cos(2\theta_{\boldalpha \d})$. The sum of\refeq{coscarre} \&\refeq{sincarre}  over $\d\in(0,r-1)^2$ yield\refeq{sumcos2_app} \&\refeq{sumsin2_app}.

\subsection{Expectations $\E[G_\boldalpha]$}

Taking the expectation of\refeq{def_Galpha} with respect to $\bop_{\d j}$ yields:
\begin{equation}
	\E G_{\boldalpha}(\k')=\sum_{\d} e^{-i\frac{2\pi}{r}(\boldalpha-\boldgamma)\cdot\d}  \E \left[e^{-i\frac{2\pi}{rN}\k'_\boldalpha\cdot \bop_{\d j}}\right]
\end{equation}
Then let $\chi(\k') =  \E \left[e^{-i\frac{2\pi}{rN}\k'\cdot \bop_{\d j}}\right]$  the characteristic function of the distribution of $\bop_{\d j}$.
It results from properties of roots of unity above that
\begin{equation}
	\sum_{\d} e^{-i\frac{2\pi}{r}(\boldalpha-\boldgamma)\d} = 
	\left\{
	\begin{array}{l}
		0 \mbox{ when } \boldalpha\neq \boldgamma,\\ 
		r^2 \mbox{ when } \boldalpha =  \boldgamma 
	\end{array}
	\right.
\end{equation}
so that denoting Kronecker's symbol by $\delta_{\boldgamma\boldalpha}$:
\begin{equation}
\label{EspG}
	\E G_{\boldalpha}(\k')=  \delta_{\boldgamma\boldalpha} \;  \chi(\k') 
\end{equation}

\subsection{The $G_\boldalpha$ are uncorrelated}
\label{Guncor}

We deal with the correlations between $G_{\boldalpha_1}(\k')$ and $G_{\boldalpha_2}(\k')$ for $\boldalpha_i\neq\boldgamma$:
\begin{multline*}
	\E [G_{\boldalpha_1}G_{\boldalpha_2}^*] 
	=  \frac{1}{n_\d^2}
	\sum_{\d\d'} e^{-i\frac{2\pi}{rN}(\boldalpha_1-\boldgamma)\d N} e^{+i\frac{2\pi}{rN}(\boldalpha_2-\boldgamma)\d' N}\\ 
	\times \sum_{j,\ell=1}^{n_\d}  
	\underbrace{\E \left[
	e^{+i\frac{2\pi}{rN}(\k'+\boldalpha_1 N)\cdot \bop_{\d j}}
	e^{-i\frac{2\pi}{rN}(\k'+\boldalpha_2 N)\cdot \bop_{\d \ell}}
	\right]}_{\beta_{j\ell}}\\
\end{multline*}
One remarks that 
\begin{equation}
\label{betajl}
	\beta_{j\ell} = 
	\left\{
	\begin{array}{cl}
		\chi((\boldalpha_2-\boldalpha_1)N) & \mbox{ if } j=\ell,\\[3mm]
		\chi(\k'_{\boldalpha_1})\chi(-\k'_{\boldalpha_2}) & \mbox{ if } j\neq \ell,\\		
	\end{array}
	\right.
\end{equation}
so that 
\begin{multline}
	\E [G_{\boldalpha_1}(\k')G_{\boldalpha_2}^*(\k')] 
	=
	 \frac{1}{n_\d^2}
	\left(\sum_{\d} e^{-i\frac{2\pi}{rN}(\boldalpha_1-\boldalpha_2)\d N}\right)\\
	\times
	\left(	\sum_{j,\ell=1}^{n_\d} \beta_{j\ell}
	- \sum_{j,\ell=1}^{n_\d} \chi(\k'_{\boldalpha_1})\chi(-\k'_{\boldalpha_2})\right)
\end{multline}
Then using\refeq{betajl} and little algebra one gets 
\begin{multline}
	\sum_{j,\ell=1}^{n_\d} \beta_{j\ell}
	- \sum_{j,\ell=1}^{n_\d} \chi(\k'_{\boldalpha_1})\chi(-\k'_{\boldalpha_2})\\
	 =
	n_\d \left[ \chi((\boldalpha_2-\boldalpha_1)N)
	- \chi(\k'_{\boldalpha_1})\chi(-\k'_{\boldalpha_2})	\right]
\end{multline}
As a consequence one finally gets:
\begin{multline}
	\E [G_{\boldalpha_1}(\k')G_{\boldalpha_2}^*(\k')] \\
	 = \delta_{\boldalpha_1\boldalpha_2} \frac{r^2}{n_\d}
	\left[ \chi((\boldalpha_2-\boldalpha_1)N)
	- \chi(\k'_{\boldalpha_1})\chi(-\k'_{\boldalpha_2})	\right] \\
	= \delta_{\boldalpha_1\boldalpha_2} \: \frac{r^2}{n_\d} \left(  1 - |\chi(\k'_{\boldalpha_1})|^2 \right) 
	\label{Guncorrelated}
\end{multline}
%
so that the $G_{\boldalpha_i}$, $\boldalpha_i\neq\boldgamma$, are uncorrelated (not independent).

\hfill QED.



\bibliographystyle{ieeetr}



\end{document}